\documentclass[twocolumn,twocolappendix,tighten,times,astrosymb]{aastex7}

\usepackage{mathtools}
\usepackage[T1]{fontenc}
\usepackage{microtype}

\newcommand{\vp}{v_{\text{push}}}         

\newcommand\bb[1]{\mbox{\boldmath{$#1$}}}

\newcommand\bcdot{\,\bb{\cdot}\,}

\newcommand\btimes{\,\bb{\times}\,}
\newcommand\rmd{\mathrm{d}}

\usepackage[dvipsnames]{xcolor}

\shorttitle{Granier et al.}
\shortauthors{Granier et al.}

\begin{document}

\title{\large 3D Kinetic Simulations of Driven Reconnection in Merging Flux Tubes}

\author[0000-0003-2841-8153]{Camille Granier}
\affiliation{Department of Physics, University of Maryland, 7901 Regents Drive, College Park, MD 20742, USA}
\affiliation{Canadian Institute for Theoretical Astrophysics, 60 St. George St, Toronto, ON M5S 3H8, Canada}
\email{camillegranier@hotmail.fr}

\author[0000-0002-7526-8154]{Fabio Bacchini}
\affiliation{Centre for mathematical Plasma Astrophysics, Department of Mathematics, KU Leuven,\\ Celestijnenlaan 200B, B-3001 Leuven, Belgium}
\affiliation{Royal Belgian Institute for Space Aeronomy, Solar-Terrestrial Centre of Excellence,\\ Ringlaan 3, 1180 Uccle, Belgium}
\email{fabio.bacchini@kuleuven.be}

\author[0000-0002-5408-3046]{Daniel Grošelj}
\affiliation{Centre for mathematical Plasma Astrophysics, Department of Mathematics, KU Leuven,\\ Celestijnenlaan 200B, B-3001 Leuven, Belgium}
\email{}

\author[0000-0002-1227-2754]{Lorenzo Sironi}
\affiliation{Department of Astronomy, Columbia University and Columbia Astrophysics Laboratory, New York, NY 10027, USA}
\affiliation{Center for Computational Astrophysics, Flatiron Institute, 162 Fifth Ave, New York, NY 10010, USA}
\email{}


\begin{abstract}
We present 2D and 3D Particle-in-Cell simulations of driven collisionless magnetic reconnection triggered by the compression and merger of two Lundquist-type force-free flux tubes in a strongly magnetized pair plasma, with a focus on magnetic energy dissipation and particle acceleration. We show that 3D effects systematically delay the onset of reconnection in comparison with equivalent 2D runs, an effect further enhanced by a strong guide field, due to reduced linear growth rates and phase decoherence of oblique modes. Increasing the external drive accelerates both tearing and drift–kink instabilities, while a strong guide field suppresses coherent drift–kink activity and has a comparatively mild impact on tearing. Despite these differences in early-time dynamics, all simulations enter a fast-merging phase characterized by a normalized reconnection rate $\sim 0.08$--$0.10$, coinciding with a transient reduction of the guide-to-reconnecting field ratio inside the current sheet. The high-energy cutoff of accelerated particles converges to a common asymptotic value, $\gamma_{\rm cut}/\sigma_{\rm in} \simeq 50$, with only a weak dependence on the driving strength. This behavior is consistent with an electric-field-limited acceleration process, in which the maximum energy is set by the reconnection electric field and the duration of the energization phase. The resulting nonthermal particle spectra are similar across all runs, with power-law indices $p \simeq 1.6$--$2.0$.
\end{abstract}

\section{Introduction}

Magnetic reconnection is a universal process that converts magnetic energy to particle energy in laboratory, heliospheric, and astrophysical plasmas \citep[see e.g., the reviews by][]{ZweibelYamada2009, YamadaKulsrudJi2010, JiDaughton2011, Sironi2025}. In high-energy scenarios, reconnection is invoked to account for rapid dissipation and particle acceleration in magnetically dominated environments, such as pulsar-wind nebulae, magnetar flares, black-hole coronae, and relativistic jets, where nonthermal emission is ubiquitously observed \citep[e.g.,][]{Uzdensky2011, Cerutti2013, GuoEtAl2015, SironiBeloborodov2019, Mehlhaff2020, WernerUzdensky2021}.

Reconnection has been extensively studied with two-dimensional (2D) models, typically in scenarios with a preexisting planar current sheet in the absence of sustained external driving \citep[e.g.,][]{Loureiro2007, Lapenta2008, TenBarge2013, Comisso2016, Huang2017, Granier2022, Granier2024}. These studies have provided foundational insights into reconnection rates, plasmoid formation, and particle acceleration mechanisms \citep{Sironi2014, Liu2015,  Hakobyan2023b}, including Fermi-like energization in contracting magnetic islands in reconnecting current sheets (see e.g.\ \citealt{Zenitani2005, Zenitani2007, BesshoBhattacharjee2012} for early kinetic studies of particle acceleration in tearing-unstable current sheets). However, real astrophysical magnetic structures are inherently 3D, often involving dynamically evolving magnetic structures such as flux-ropes \citep[e.g.,][]{Lyutikov2017, Ripperda2017a}. 3D effects introduce additional complexities such as oblique tearing modes, drift-kink modes, and nonplanar current-sheet deformations, which can fundamentally alter the reconnection process in comparison with 2D \citep{Borgogno2005, Zenitani2005, Zenitani2008, Daughton2011,  Cerutti2014, Guo2015, Stanier2015, Dahlin2017, Guo2021, Hakobyan2021, WernerUzdensky2021, Zhang2021, Hoshino2024, Bacchini2025}. In particular, fully 3D reconnection permits particle escape from magnetic islands or flux ropes, enabling fast, sustained energization that is suppressed in strictly 2D geometries \citep[e.g.,][]{Dahlin2015, Dahlin2017, Petropoulou2018, Zhang2021, Zhang2023}. 

In this paper, we study 3D \emph{driven} magnetic reconnection in a collisionless electron--positron plasma, building on our previous 2D flux-tube-merging study \citep{Granier2025}. The paper is organized as follows. Section~\ref{sec:setup} describes the numerical setup. Section~\ref{sec:tearing_kink} analyzes current-sheet formation, instability growth, and reconnection rates. Particle acceleration and energy spectra are discussed in Section~\ref{sec:accel_mechanism}. Our conclusions are summarized in Section~\ref{sec:conclusion}.

\section{Numerical Setup}
\label{sec:setup}

We perform 3D particle-in-cell (PIC) simulations using the \textsc{Tristan-MP v2} code \citep{Hakobyan2023}. Our numerical setup follows that of our previous 2D study \citep{Granier2025}.

\begin{figure*}[ht!]
\centering
\includegraphics[width=1\textwidth]{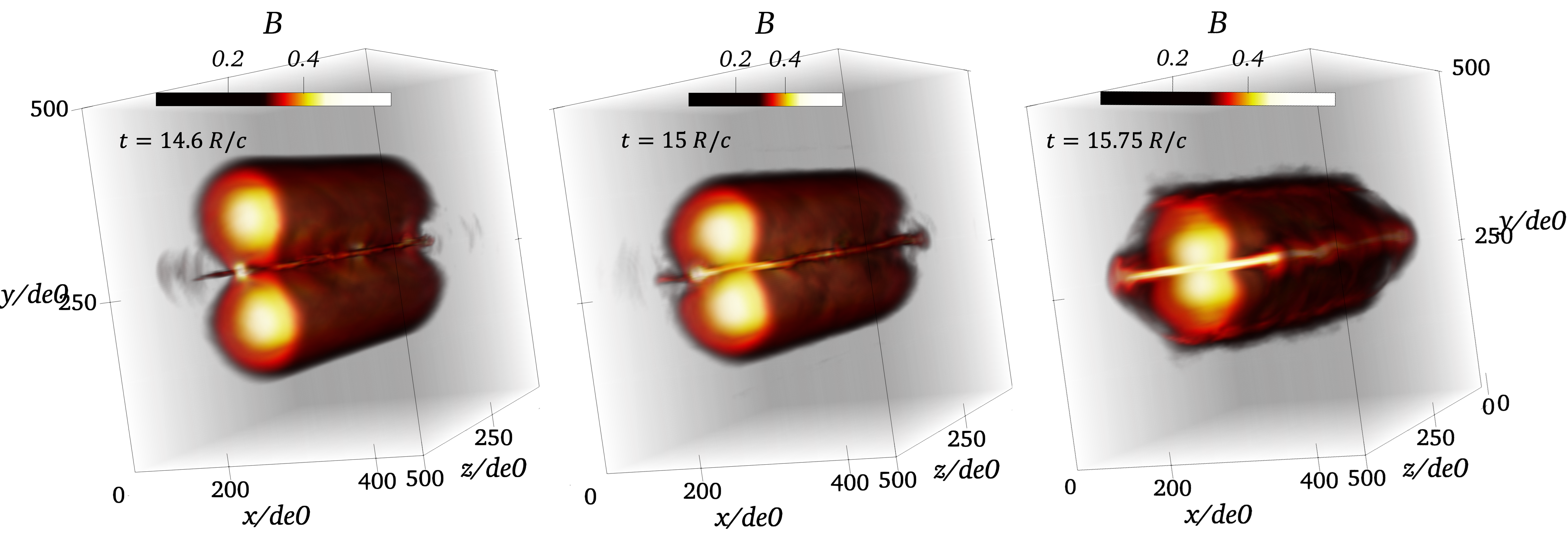}
\caption{3D visualization of the magnetic-field amplitude for the case with $v_{\mathrm{push}} = 0.02c$ and $C = 10^{-4}$ at subsequent times (from left to right) during the merging.}
\label{fig:3Dviz}
\end{figure*}

We initialize the system with two identical cylindrical Lundquist-type force-free flux tubes, each with radius $R$, embedded in a uniform electron--positron plasma of total density $n_0$. The initial magnetic-field configuration within each flux tube is defined by the magnetic field \citep{Lyutikov2017, Ripperda2017a}
\begin{equation}
B_\phi(r) = B_0 J_1(\alpha r/R),
\end{equation}
\begin{equation}
B_z(r) = B_0 \sqrt{J_0^2(\alpha r/R) + C},
\end{equation}
where $r$ is the radial distance from the tube axis, $J_0$ and $J_1$ are Bessel functions of the first kind, $\alpha \approx 3.8317$ is the first zero of $J_1$, $B_0$ sets the characteristic magnetic-field strength, and the constant $C$ controls the relative guide-field contribution in the vicinity of the current-sheet midplane where reconnection occurs.
The tubes are placed in a cubic domain of side $4R$ with tube axes located at $(x,y,z)=(2R,R,2R)$ and $(2R,3R,2R)$ and periodic boundaries. To seed reconnection, we impose on each tube an equal and opposite drift-velocity push $\bb{v}_{\rm push}=\mp\,v_{\rm push}\,\hat{\boldsymbol y}$ for the two tubes. This induces an initial motional electric field $\bb{E}=-(\bb{v}_{\rm push}\btimes\bb{B})/c$ which drives the tubes toward the midplane at $y=2R$. Importantly, we do not impose any additional perturbation: the initial condition is invariant under translations along $z$ (up to particle noise and numerical roundoff), so all 3D structures develop self-consistently from the growth of 3D instabilities. 
Fig.~\ref{fig:3Dviz} shows a representative 3D rendering of the flux tubes during the merging phase.

All 3D simulations are performed on a grid with $1600^3$ cells. The plasma skin depth is resolved with three cells ($d_{e0}=c/\sqrt{4\pi n_0 e^2/m_e}=3\,\Delta x$), and each cell initially contains 28 particles per species. The background plasma is cold, with $kT_0=0.005\,m_e c^2$.

We adopt a total magnetization $\sigma_0 = B_0^2/(4\pi n_0 m_e c^2)=40$. The in-plane magnetization, computed from the reconnecting magnetic field in the $(x,y)$-plane, is $\sigma_{\rm in}=0.16\,\sigma_0=6.4$. We consider four 3D simulations: three runs at fixed $C=10^{-4}$ with drive strengths $v_{\rm push}/c=\{0.02,\,0.1,\,0.6\}$, and one additional run at $v_{\rm push}=0.1c$ with a stronger guide field, $C=0.2$. The role of the control parameter $C$, which tunes the relative guide-field strength and slightly perturbs exact force-free balance, is discussed in detail in Appendix~\ref{app:cparam}. For direct comparison, we also perform corresponding 2D simulations with identical parameters.

To characterize the effective guide-field strength during reconnection, we define $B_g$ as the $B_z$ component evaluated inside the current sheet and $B_{\rm up}$ as the reconnecting in-plane field immediately upstream of the current sheet at the location where the sheet thickness is measured. The ratio $B_g/B_{\rm up}$ is computed locally and averaged in time over the quasi-steady reconnection phase defined as the interval between the onset of reconnection and the beginning of the fast merging; its temporal evolution is presented in Section~\ref{sec:rate}. For the three runs with $C=10^{-4}$, this procedure yields $\langle B_g/B_{\rm up}\rangle \simeq 0.7$, while the run with $C=0.2$ yields $\langle B_g/B_{\rm up}\rangle \simeq 1.3$.

As shown in \citet{Granier2025}, we distinguish three stages in the evolution. \emph{Phase I:} the current sheet forms; \emph{Phase II:} reconnection has started but proceeds very slowly, until $\sim 1\%$ of the in-plane magnetic energy, associated with the reconnecting field, is depleted; \emph{Phase III:} the tubes coalesce rapidly, and we define $t_{\rm merg}$ as the moment when the in-plane magnetic energy contained in the flux tubes decreases by $1\%$, $t_{\rm relax}$ at $30\%$, and $\Delta t_{\rm merg}\equiv t_{\rm relax}-t_{\rm merg}$.

\section{Tearing vs.\ Drift-Kink}
\label{sec:tearing_kink}

\subsection{Current-sheet Formation and Collapse: 2D vs.\ 3D}

\begin{figure}
    \centering
    \includegraphics[width=1\linewidth]{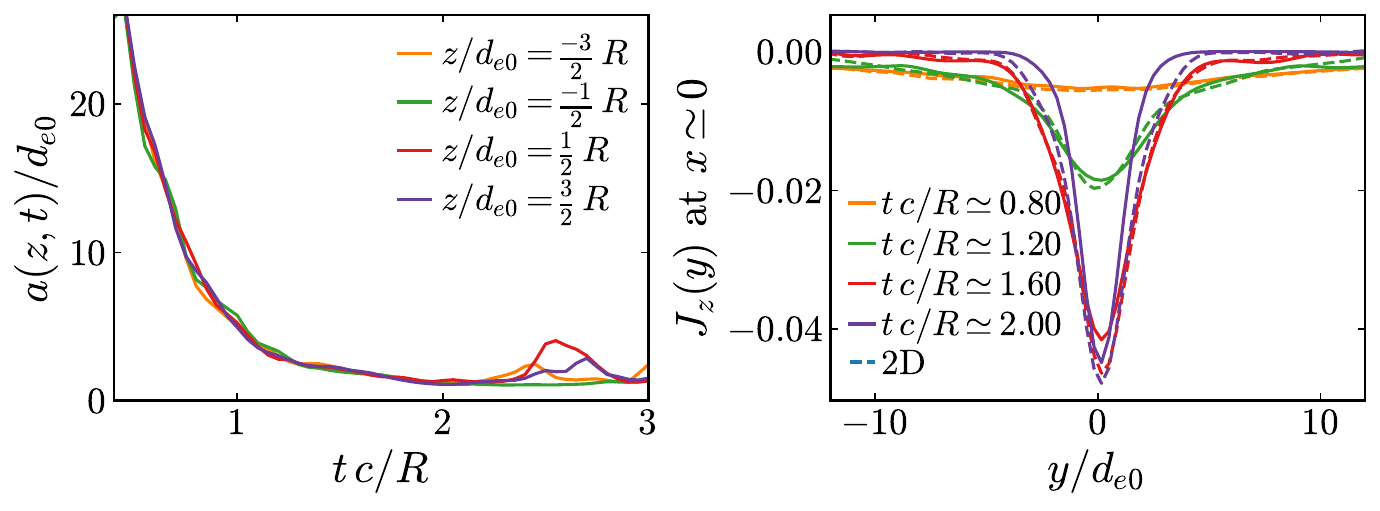}
    \caption{Left: Time evolution of the current-sheet half-thickness $a/d_{e0}$ measured at several $z$-planes. Right: transverse cuts $J_z(y)$ at $x\simeq 0$. 3D profiles are averaged over all $z$-slices (solid). The corresponding 2D profiles are shown with dashed lines. For this run, $v_{\rm push}=0.6c$. }
    \label{fig:widthcuts}
\end{figure}

The left panel of Fig.~\ref{fig:widthcuts} shows the time evolution of the current-sheet half-thickness, $a(z,t)$, measured at several fixed $z$-planes across the domain and at $x\simeq 0$, i.e.\ through the center of the reconnecting layer. For each fixed $z$, the half-thickness is defined from the transverse profile $J_z(y)$ as the distance from the current maximum to the point where $J_z$ drops to two-thirds of its peak value, using a local polynomial fit around the peak. During the entire thinning phase, the curves at different $z$ are nearly indistinguishable, implying that the large-scale thinning remains close to translation-invariant along $z$. The right panel compares $J_z(y)$ cuts at matched physical times between 2D and 3D. In the 3D case, the solid curve is additionally averaged over all $z$-slices to highlight the global sheet structure; this $z$-averaged profile closely tracks the corresponding 2D cut over the same $y$ window. Thus, up to the end of the thinning stage, the 3D evolution remains effectively quasi-2D when characterized by $z$-averaged quantities.
This trend is observed for all simulations. 

\begin{figure}
    \centering
    \includegraphics[width=1\linewidth]{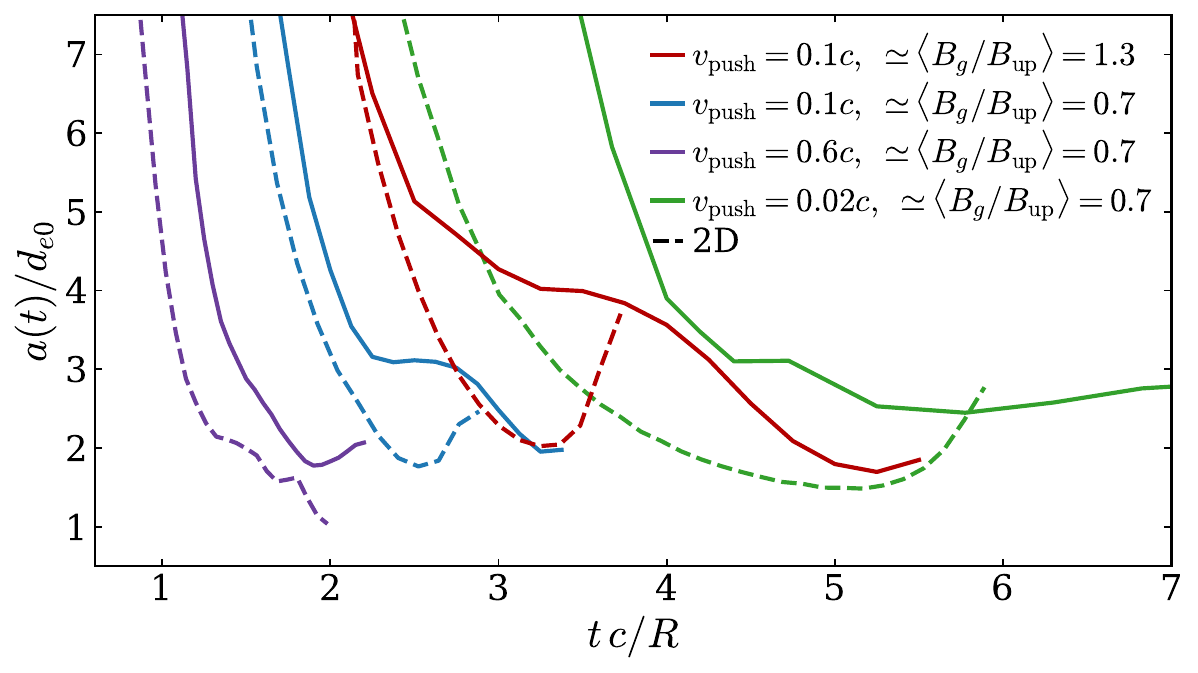}
    \caption{Time evolution of the current-sheet half-thickness $a/d_{e0}$, for both 2D (dashed lines) and 3D (solid lines) simulations with different driving velocities $v_{\rm push}$ and guide-field strengths. Later times, when the thickness starts increasing due to magnetic islands growing, are not shown.}
    \label{fig:width_vs_time}
\end{figure}

Fig.~\ref{fig:width_vs_time} shows the sheet-width evolution for all runs. For the 3D runs, after an initial exponential thinning similar to 2D, the width temporarily increases or stalls before resuming its decrease. This is more pronounced for the case  $\langle B_g/B_{\rm up}\rangle \simeq 1.3$. 
We interpret this trend as the combination of several effects: (i) \emph{Oblique tearing:} A finite guide field naturally favors oblique tearing modes in 3D \citep{Kuznetsova1985}. Unstable perturbations satisfy the resonance condition $\boldsymbol{k}\bcdot\boldsymbol{B}(y_s)=0$, which defines resonant surfaces at fixed $y=y_s$ across the sheet. In practice, different $(k_x,k_z)$ modes resonate at different transverse positions, so the tearing activity is distributed over slightly different layers within the sheet. As a consequence, different $z$-planes reconnect at slightly tilted orientations and out of phase, producing a patchy current sheet whose $z$-averaged thickness can temporarily stall before global thinning resumes \citep[e.g.,][]{Daughton2011}. To illustrate the geometric effect of obliquity, consider a Harris-type profile $B_{Hx}=B_{H,\mathrm{up}}\tanh(y/a)$, where $B_{H,\mathrm{up}}$ is the reconnecting component of the illustrative Harris equilibrium. The resonance condition then implies $\tanh(y_s/a)=-r$, with $r\equiv(k_z/k_x)(B_{Hg}/B_{H,\mathrm{up}})$, which reduces the local magnetic shear to $(\mathrm{d}B_{Hx}/\mathrm{d}y)_{y_s}=(B_{H,\mathrm{up}}/a)(1-r^2)$. Oblique modes therefore behave as if they were embedded in an effectively thicker layer of width $a_{\rm eff}\sim a/(1-r^2)$. While this construction is only illustrative, it highlights that for oblique modes that remain resonant, a finite guide field reduces the local magnetic shear experienced by the instability and delays its growth. (ii) \emph{Reduced linear growth:} linear tearing growth rates decrease with increasing guide field for oblique modes, so that a larger $B_g/B_{\rm up}$ requires a longer time for perturbations to reach nonlinear amplitude \citep[e.g.,][]{Daughton2011, Baalrud2018}. 
(iii) \emph{Magnetic pressure:} a finite guide  field contributes an extra pressure term to the cross-sheet force balance, so that as $B_g$ is advected into the layer, the additional magnetic pressure resists compression and reduces the thinning rate. 
(iv) \emph{Competition with drift--kink:} in the weak-guide-field limit, thin current sheets in 3D pair plasmas are susceptible to the drift-kink instability, which broadens the layer. In our simulations, however, the drift-kink instability starts growing after the first signs of reconnection and is stabilized when increasing the guide-field amplitude. A direct tearing vs.\ drift-kink comparison is presented in the next section.
Taken together, these effects indicate that the guide field introduces a generic delay of reconnection onset through reduced linear growth and additional magnetic support (points ii–iii), which operates in both 2D and 3D, while the transient stalling of the width observed in our simulations arises specifically from 3D oblique tearing (point i).

\subsection{Mode Analysis}

\begin{figure}
    \centering
    \includegraphics[width=1\linewidth]{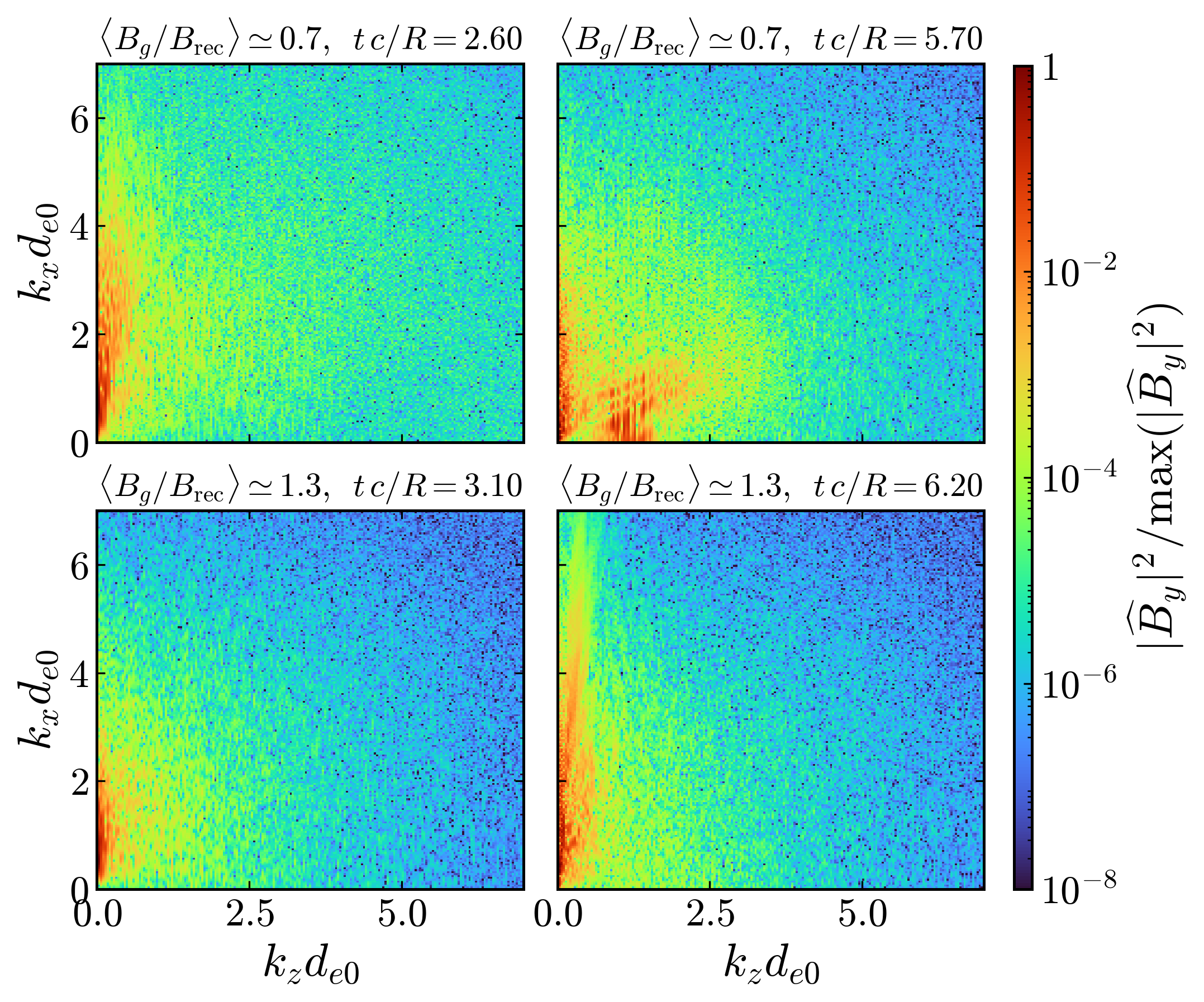}
    \caption{2D power spectrum $|\widehat{B_y}(k_x,k_z)|^2$ of the reconnecting magnetic field for weak ($\langle B_g/B_{\rm up}\rangle\simeq0.7$, top) and strong ($\langle B_g/B_{\rm up}\rangle\simeq1.3$, bottom) guide-field cases at subsequent times (from left to right). Increasing the guide field suppresses oblique power (large $k_z$), concentrating energy at small $k_z$ and yielding a more anisotropic, quasi-two-dimensional spectrum.}
    \label{fig:By_kxkz_bg0}
\end{figure}

\begin{figure*}
    \centering
    \includegraphics[width=1.0\linewidth]{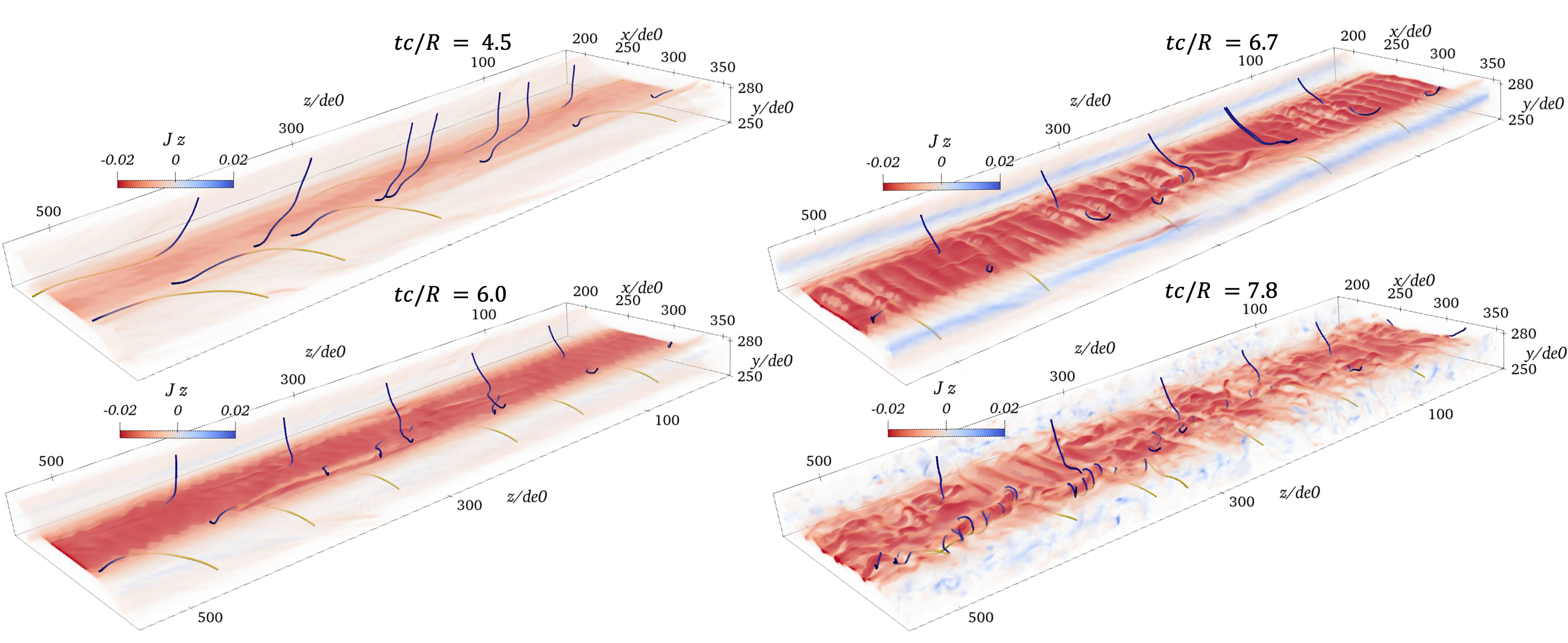}
    \caption{3D visualization of the current sheet for  $v_{\mathrm{push}} = 0.02c$ and $C = 10^{-4}$ at subsequent times. The volume rendering shows the spatial distribution of $J_z$ in the sheet, with highlighted magnetic-field lines. Between $tc/R = 4.5 $ and $tc/R = 6.0 $, localized reconnection is already visible: small flux-rope precursors form and remain coherent prior to the development of the drift-kink instability at $tc/R = 6.0 $.}
    \label{fig:3Dcd}
\end{figure*}

\begin{figure*}[t]
\centering
\includegraphics[width=0.49\textwidth]{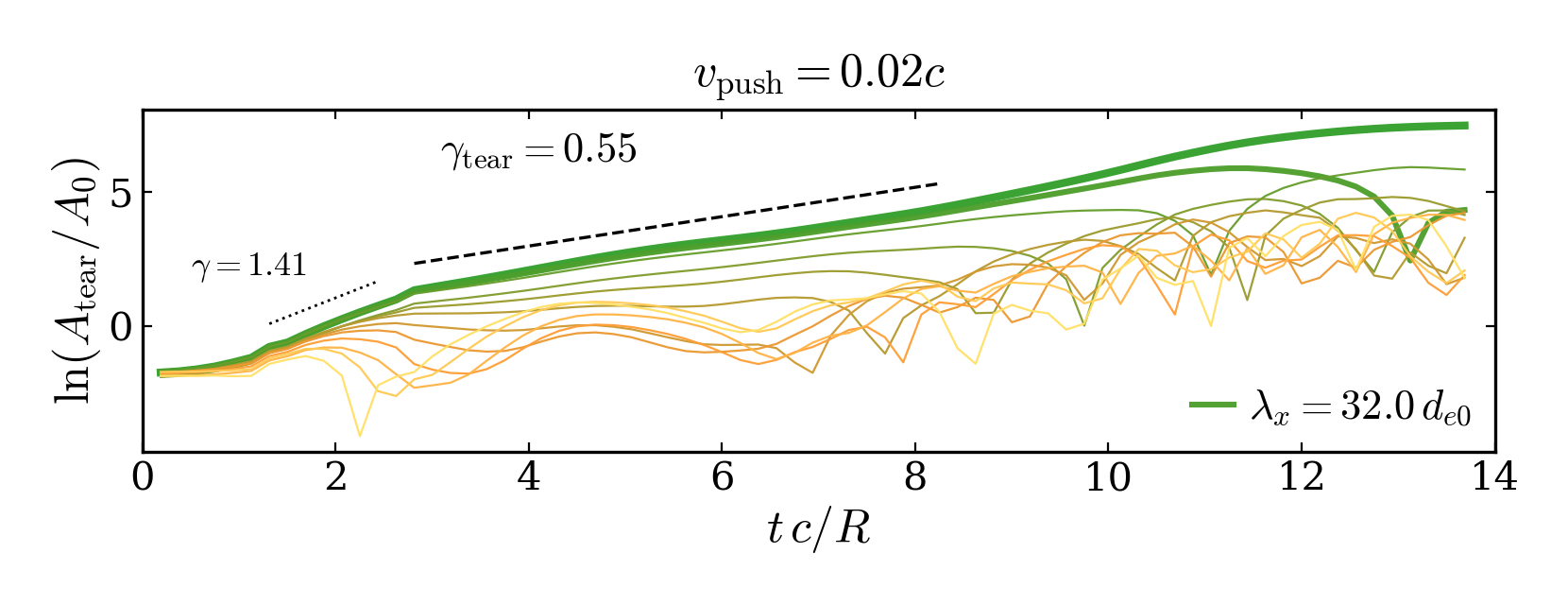}
\includegraphics[width=0.49\textwidth]{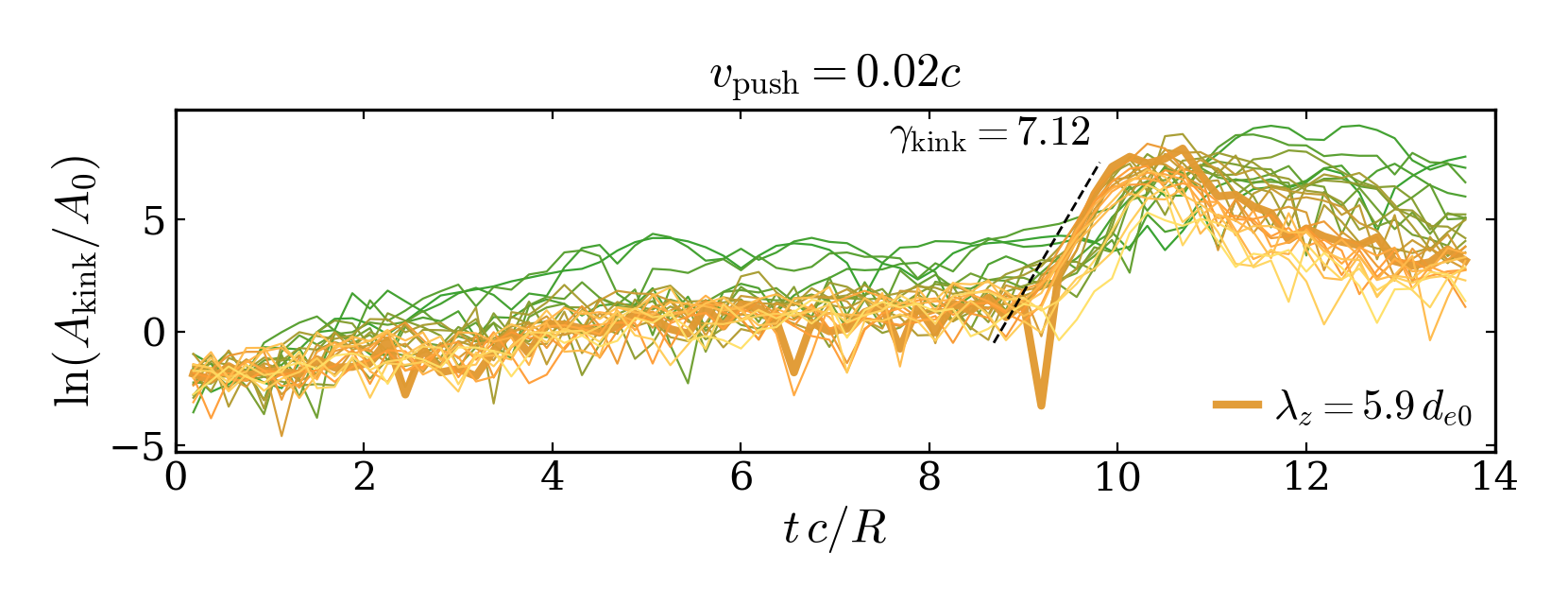}
\includegraphics[width=0.49\textwidth]{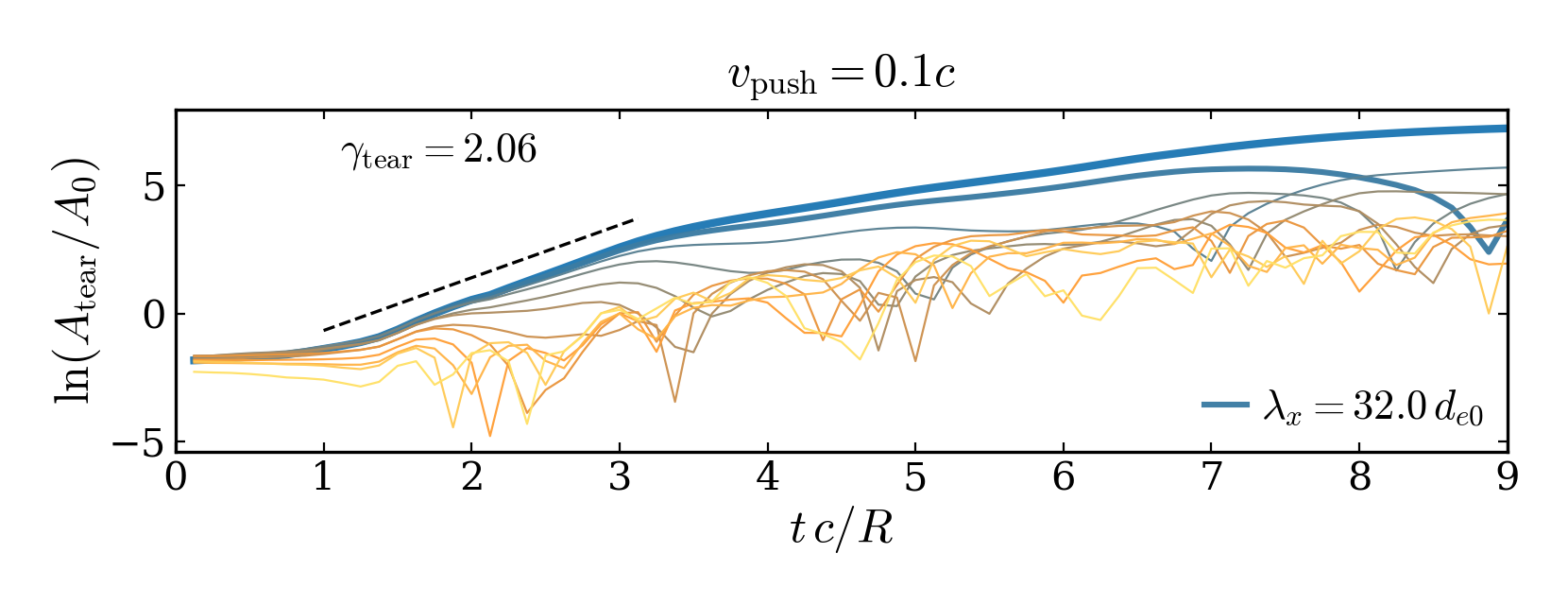}
\includegraphics[width=0.49\textwidth]{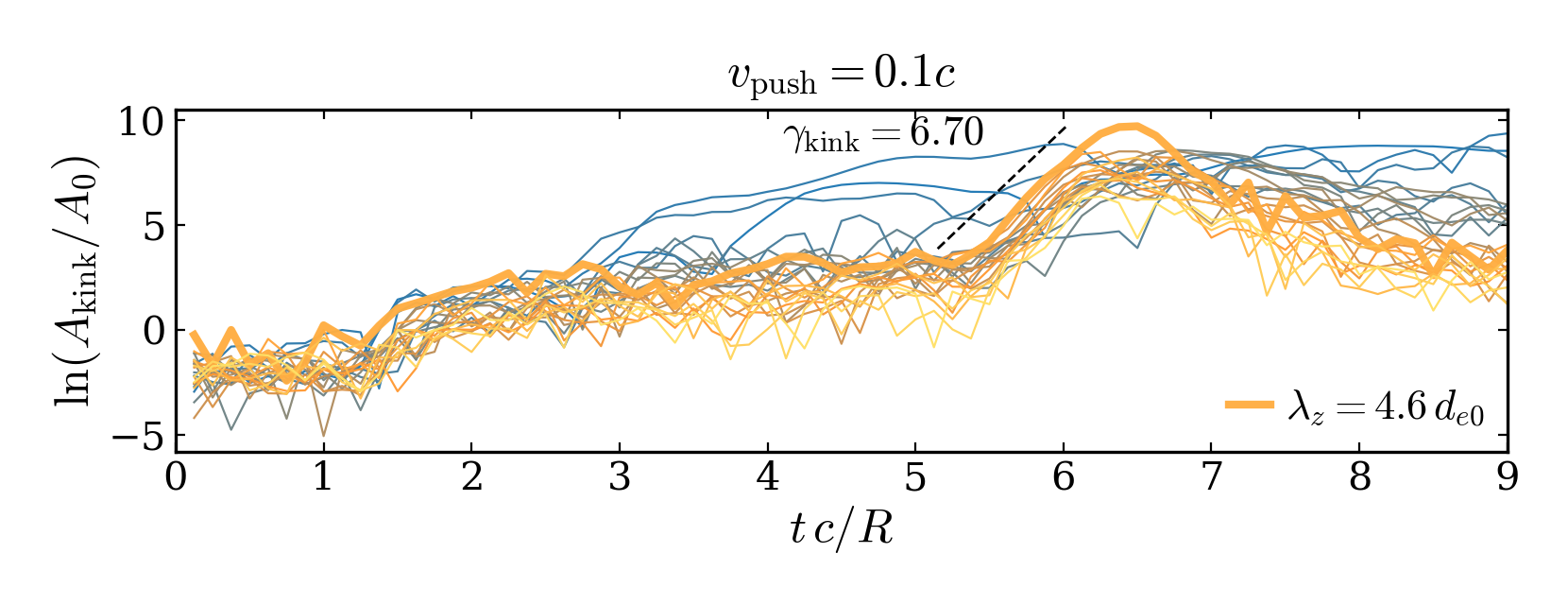}
\includegraphics[width=0.49\textwidth]{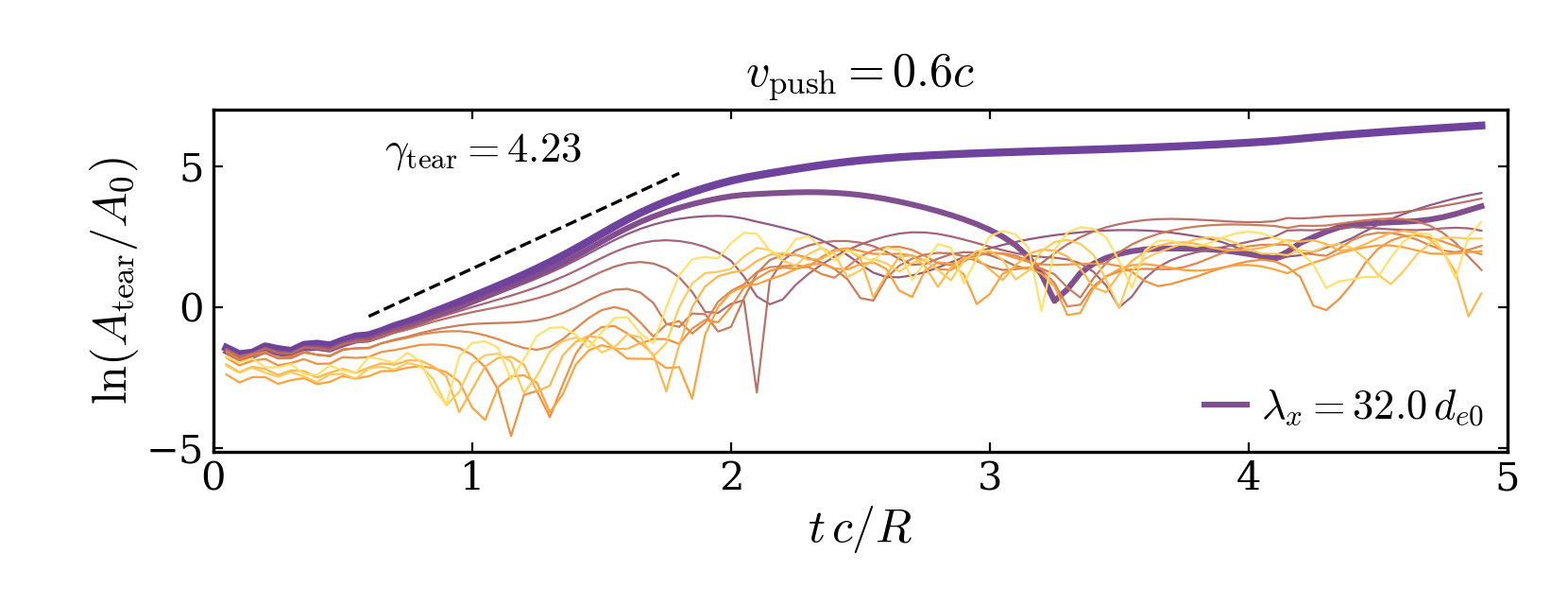}
\includegraphics[width=0.49\textwidth]{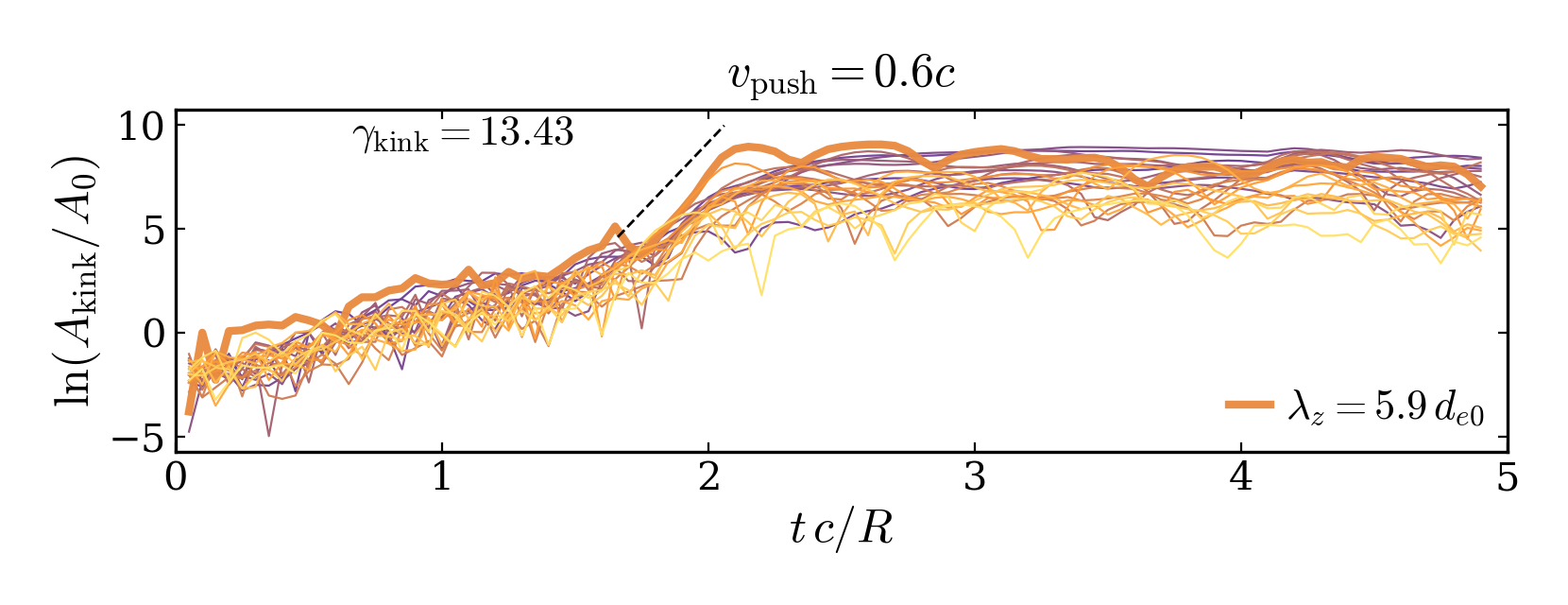}
\caption{Growth of tearing (left panels) and drift-kink (right panels) instabilities. The fitted modes are emphasized with thicker lines, and black dashed curves show exponential fits of the growth rates $\gamma_{\rm tear}$ and $\gamma_{\rm kink}$.}
\label{fig:mode_growth_vpush}
\end{figure*}

\begin{figure}[t]
\centering
\includegraphics[width=1\linewidth]{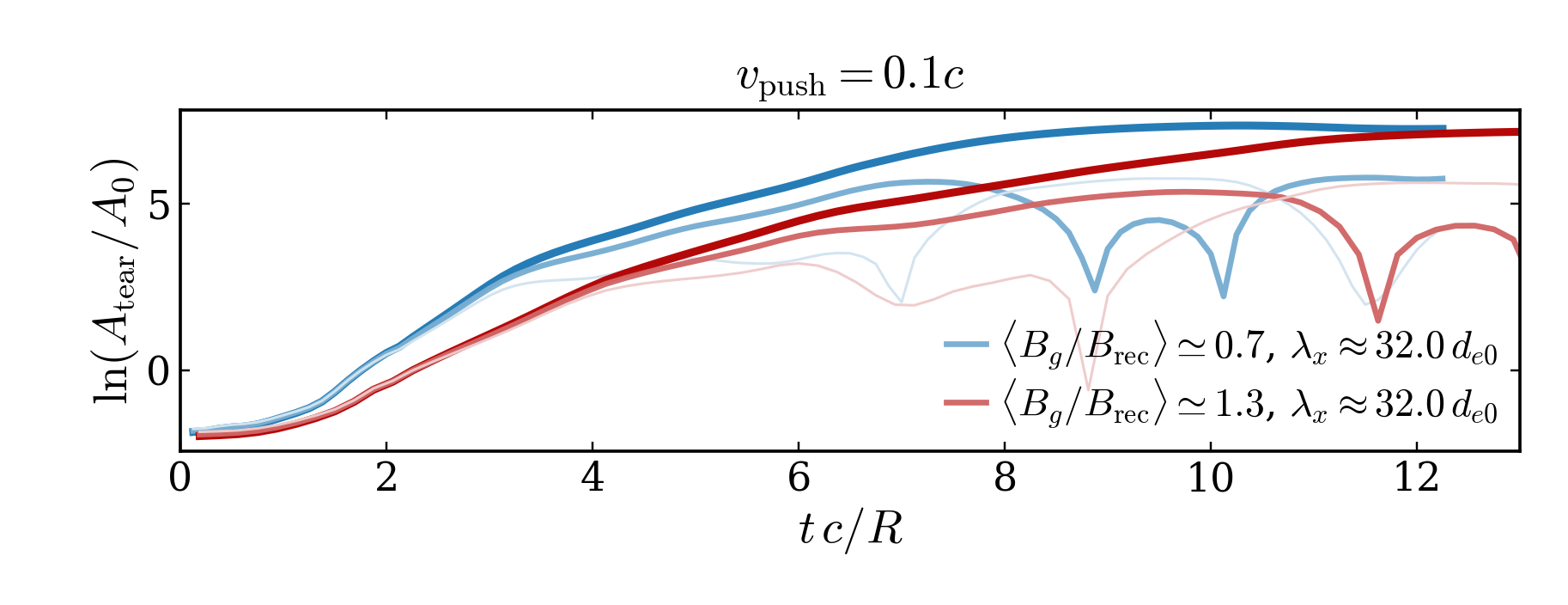}
\includegraphics[width=1\linewidth]{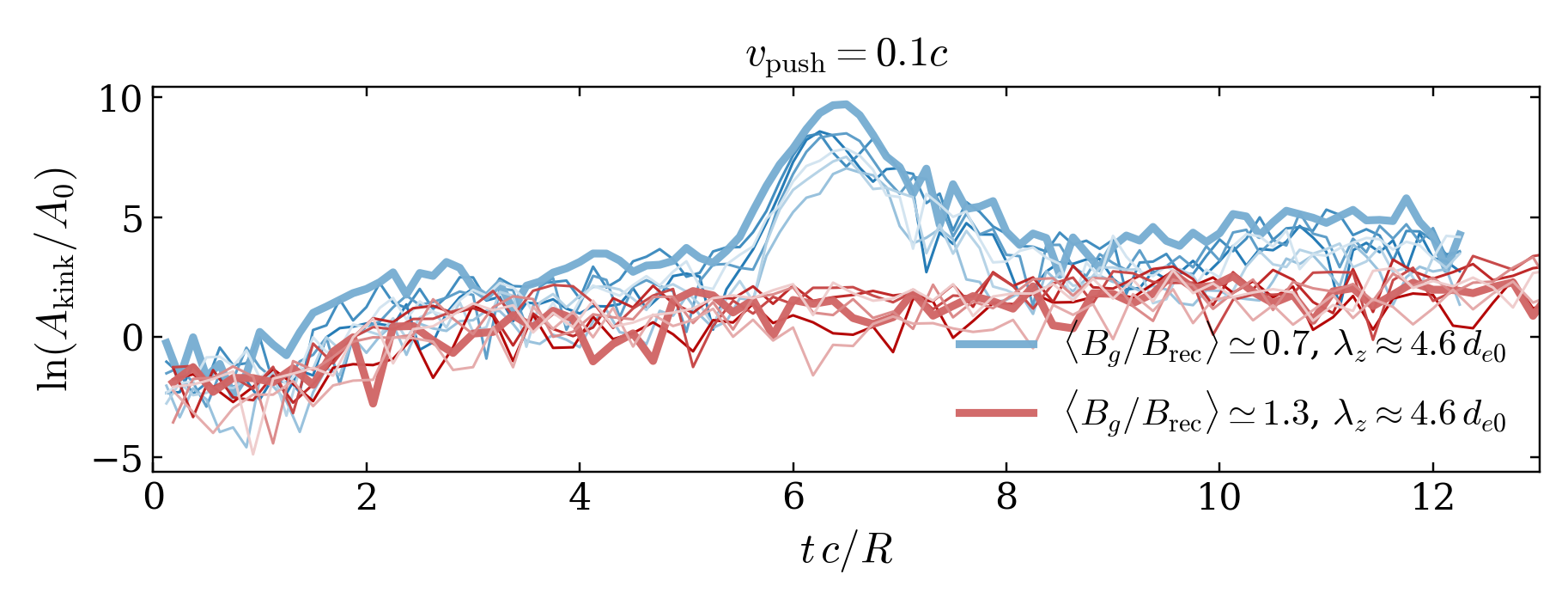}
\caption{Comparison of mode growth for weak ($\langle B_g/B_{\rm up}\rangle\simeq0.7$) and strong ($\langle B_g/B_{\rm up}\rangle\simeq1.3$) guide field at $v_{\mathrm{push}}=0.1c$.}
\label{fig:mode_growth_guidefield}
\end{figure}

Fig.~\ref{fig:By_kxkz_bg0} shows the two–dimensional power spectra of the reconnecting magnetic-field component $B_y$ in $(k_x,k_z)$ space for weak and strong guide fields with $v_{\mathrm{push}}=0.1c$. In both the initial thinning phase and the subsequent nonlinear evolution, the spectra are dominated by power at small $k_z$, indicating that most of the magnetic energy is carried by modes that vary weakly along the current direction. Increasing the guide field further suppresses oblique power, rendering the dynamics increasingly 
quasi-2D at the level of the global spectrum. Here, $m_x$ and $m_z$ denote the integer Fourier mode numbers along the $x$- and $z$-direction, respectively. For completeness, we have included in Appendix~\ref{app:fastest_modes} a comparison of the six fastest–growing $(m_x,m_z)$ mode pairs for slow and fast driving. This comparison shows that increasing the push accelerates the growth of the entire mode spectrum rather than selecting a single fastest-growing mode. A more precise analysis of which linear mode grows faster remains beyond the scope of the present simulations and will be pursued in future work.

The 3D evolution of the current sheet for the reference case $v_{\rm push}=0.1c$ with weak guide field is shown in Fig.~\ref{fig:3Dcd}. The volume rendering highlights the out-of-plane current density $J_z$.  The magnetic-field lines (blue and yellow) indicate the onset of reconnection before the development of the drift–kink instability.

We observe that, at small $(k_x,k_z)$, the power is associated with tearing-like perturbations of the reconnecting field, while enhanced power at larger $k_z$ corresponds to drift–kink–type distortions along the current direction. 
To isolate the underlying dynamics, we decompose the magnetic perturbations into tearing and drift–kink contributions and track the time evolution of their mode amplitudes (Figs.~\ref{fig:mode_growth_vpush} and \ref{fig:mode_growth_guidefield}). The tearing amplitude is defined as
$A_{\rm tear}(k_x,t)\equiv \langle|\widehat{B_y}(k_x,k_z,t)|\rangle_{k_z}$,
and the drift-kink amplitude as
$A_{\rm kink}(k_z,t)\equiv \langle|\widehat{B_x}(k_z,k_x,t)|\rangle_{k_x}$.
Linear growth rates are extracted from the intervals of steepest exponential rise of the dominant Fourier modes. For tearing, we see a clear and systematic dependence on the imposed drive:  $\gamma_{\rm tear}\simeq 0.61 c/R$ ($v_{\rm push}=0.02c$),  $2.0c/R$ ($0.1c$), and  $4.19c/R$ ($0.6c$). In all cases, the mode saturates at amplitudes of order $10^{-6}$--$10^{-8}$.  
Drift-kink also shows a strong dependence on drive strength, although not strictly monotonic over our limited set of runs. For weak guide field, we find $\gamma_{\rm kink}\simeq 7.12\,c/R$ at $v_{\rm push}=0.02c$, $6.7\,c/R$ at $0.1c$, and $13.43\,c/R$ at $0.6c$, with dominant wavelengths $\lambda_z\simeq 4.6$--$5.9\,d_{e0}$. 

The guide field primarily affects the drift-kink mode: comparing runs at $v_{\rm push}=0.1c$, increasing $\langle B_g/B_{\rm up}\rangle$ from $0.7$ to $1.3$ nearly eliminates coherent kink growth, while the tearing mode changes only mildly and saturates at similar amplitudes (consistent with the results of \citealt{Cerutti2014, Barkov2016}). Thus, tearing is always present and accelerated by the external drive, whereas the drift–kink instability operates only in weak–guide configurations and develops after tearing, where it distorts the already-formed current sheet and contributes to the late-time turbulent dynamics.

\subsection{Reconnection Rate}
\label{sec:rate}

\begin{figure}
    \centering
    \includegraphics[width=1\linewidth]{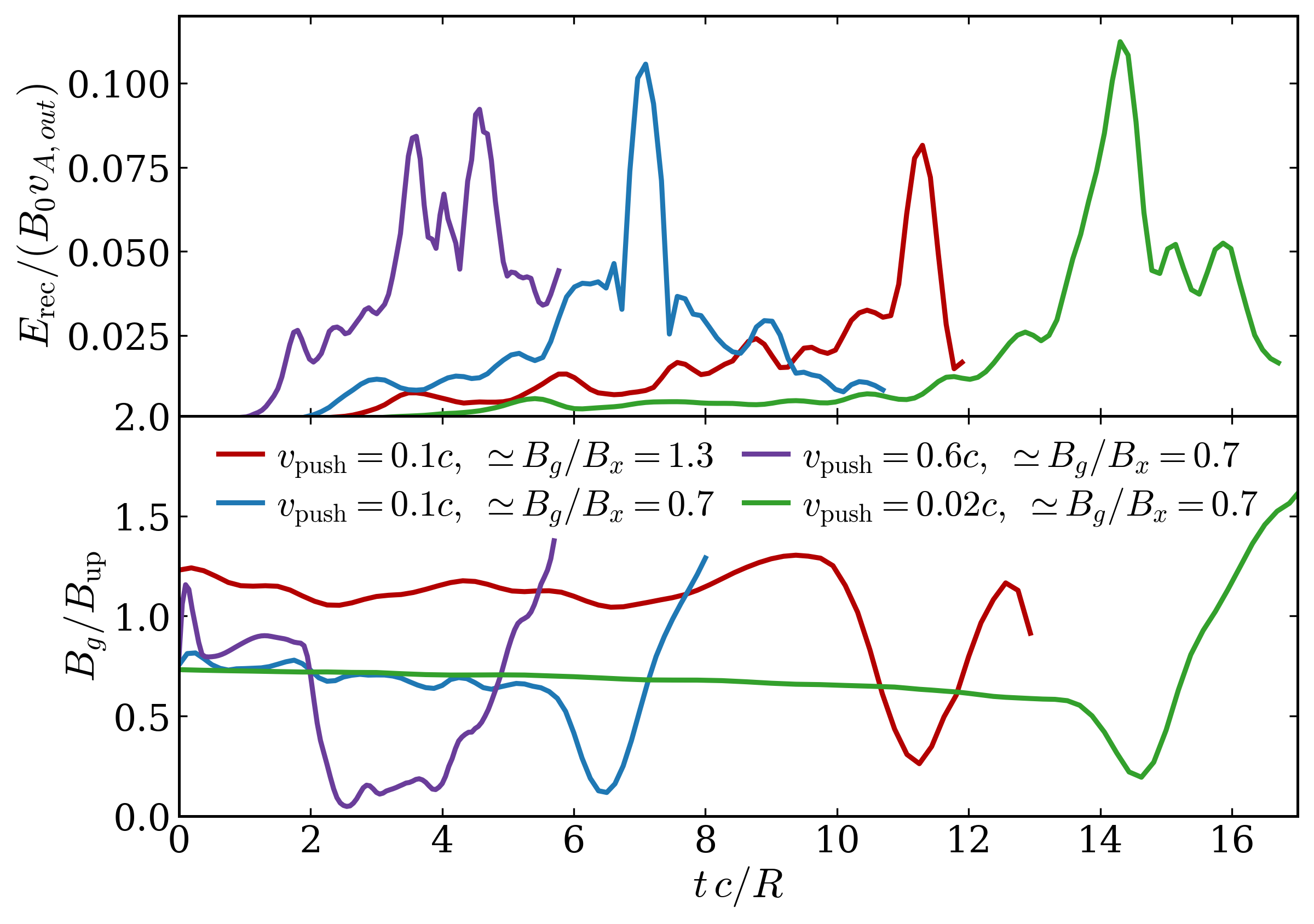}
    \caption{Time evolution of the reconnection rate (top) and the guide-field strength (bottom) for different driving velocities $v_{\rm push}$ and guide-field ratios $\langle B_g/B_{\rm up}\rangle$. $B_{\rm up}$ is measured upstream and $B_g$ is measured at the center of the current sheet.}
    \label{fig:rate}
\end{figure}

Fig.~\ref{fig:rate} shows the temporal evolution of the normalized reconnection rate $E_{\rm rec}/(B_{\rm up} v_{A,\rm out})$ (top) and the guide-to-reconnecting ratio $\langle B_g/B_{\rm up}\rangle$ (bottom) for the 3D runs. Here, $E_{\rm rec}$ is the reconnection electric field at the X-line, $v_{A,\rm out}$ is the outflow Alfv\'en speed, and $B_{\rm up}$ is the reconnecting component of the upstream magnetic field. The Alfv\'en speed is measured using the time-dependent upstream field $B_{\rm up}(t)$. $E_{\rm rec}$ is inferred from the drift speed $\langle |E_z B_y|/B^2 \rangle$ averaged over $x$--$y$ and over a narrow slab around the $z$-location of maximal inflow. In all runs, $\langle B_g/B_{\rm up}\rangle$ drops markedly during the fast-merging phase. The reconnection peaks track this drop: for weak guide field, $E_{\rm rec} c /(B_{\rm up} v_{A,\rm out})$  reaches $\sim0.05$--$0.07$ on average during merging, with peaks up to $0.09$--$0.10$; the strong-guide-field case shows a delayed start of the fast merging phase and a lower peak ($\sim0.08$), and $B_g/B_{\rm up}$ does not fall as low.
This correlation can be understood by noting that while the reconnecting field $B_{\rm up}$ provides the magnetic tension that accelerates the exhaust, the guide field $B_g$ contributes additional inertia that must be advected by the outflow. A simple estimate of the effective outflow Alfv\'en speed therefore treats $B_{\rm up}$ as the driving component and $B_g$ as an inertial load, yielding \citep{Parker1963}
\begin{equation}
v_{A,\rm out}(t) \sim c\sqrt{\frac{\sigma_{\rm up}(t)}{1+\sigma_{\rm up}(t)+\sigma_g(t)}},
\end{equation}
 where $\sigma_{\rm up}(t)\equiv B_{\rm up}^2(t)/(4\pi n_0 m_e c^2)$ and
$\sigma_g(t)\equiv B_g^2(t)/(4\pi n_0 m_e c^2)$. With $E_{\rm rec}\simeq v_{\rm in}B_{\rm up}=\epsilon\,v_{A,\rm out}B_{\rm up}$, this immediately implies that a decrease in $B_g/B_{\rm up}$ increases $v_{A,\rm out}$ and hence boosts $E_{\rm rec}$ at fixed efficiency $0.1 \equiv v_{\rm in}/v_{A,\rm out}$.

\section{Particle acceleration}
\label{sec:accel_mechanism}

\subsection{Nonthermal Particle Spectra}

\begin{figure}
    \centering
    \includegraphics[width=1\linewidth]{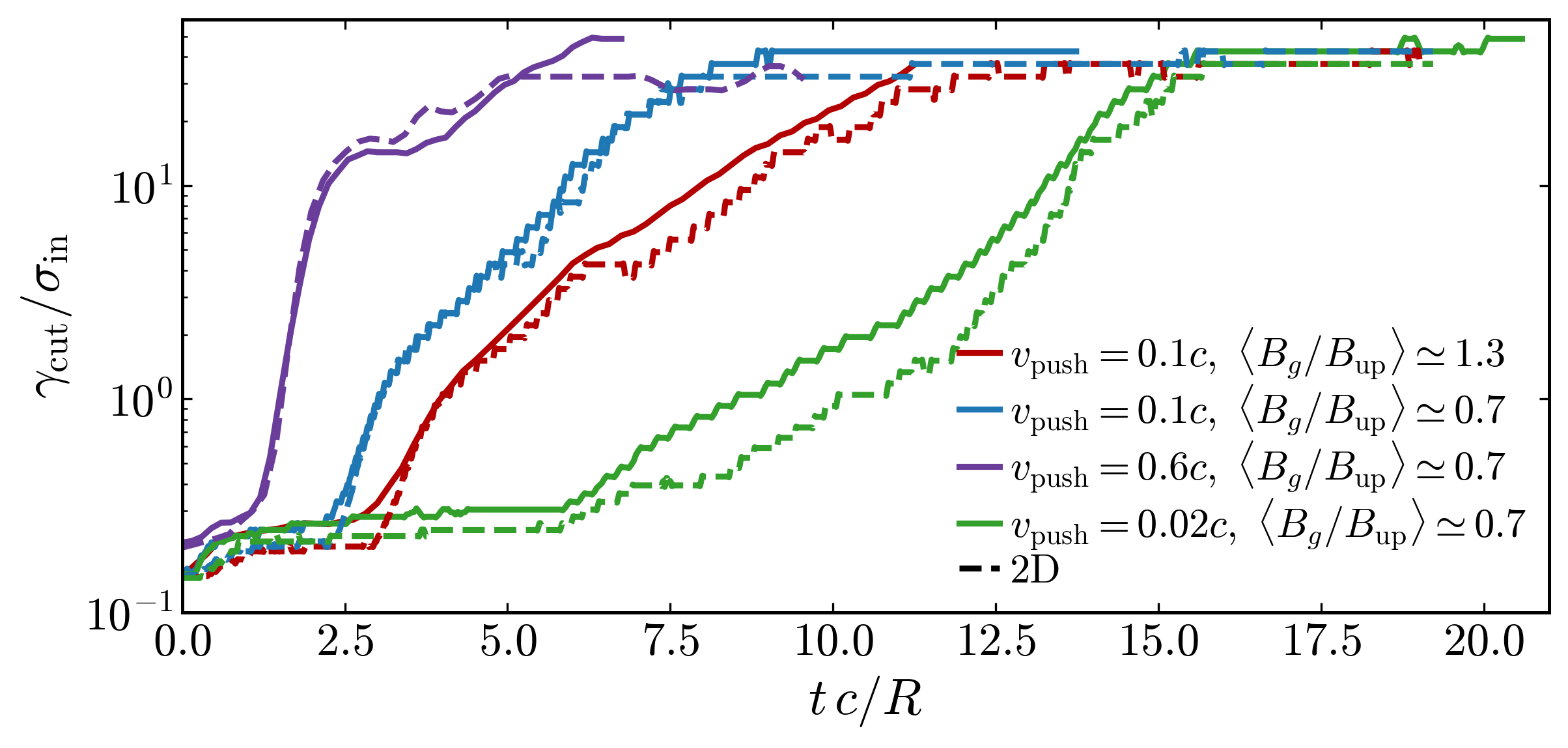}
    \caption{Time evolution of the high-energy cutoff of accelerated particles $\gamma_{\rm cut}/\sigma_{\rm in}$ 3D (solid) and 2D (dashed).}
    \label{fig:cutoff}
\end{figure}
\begin{figure}
    \centering
    \includegraphics[width=1\linewidth]{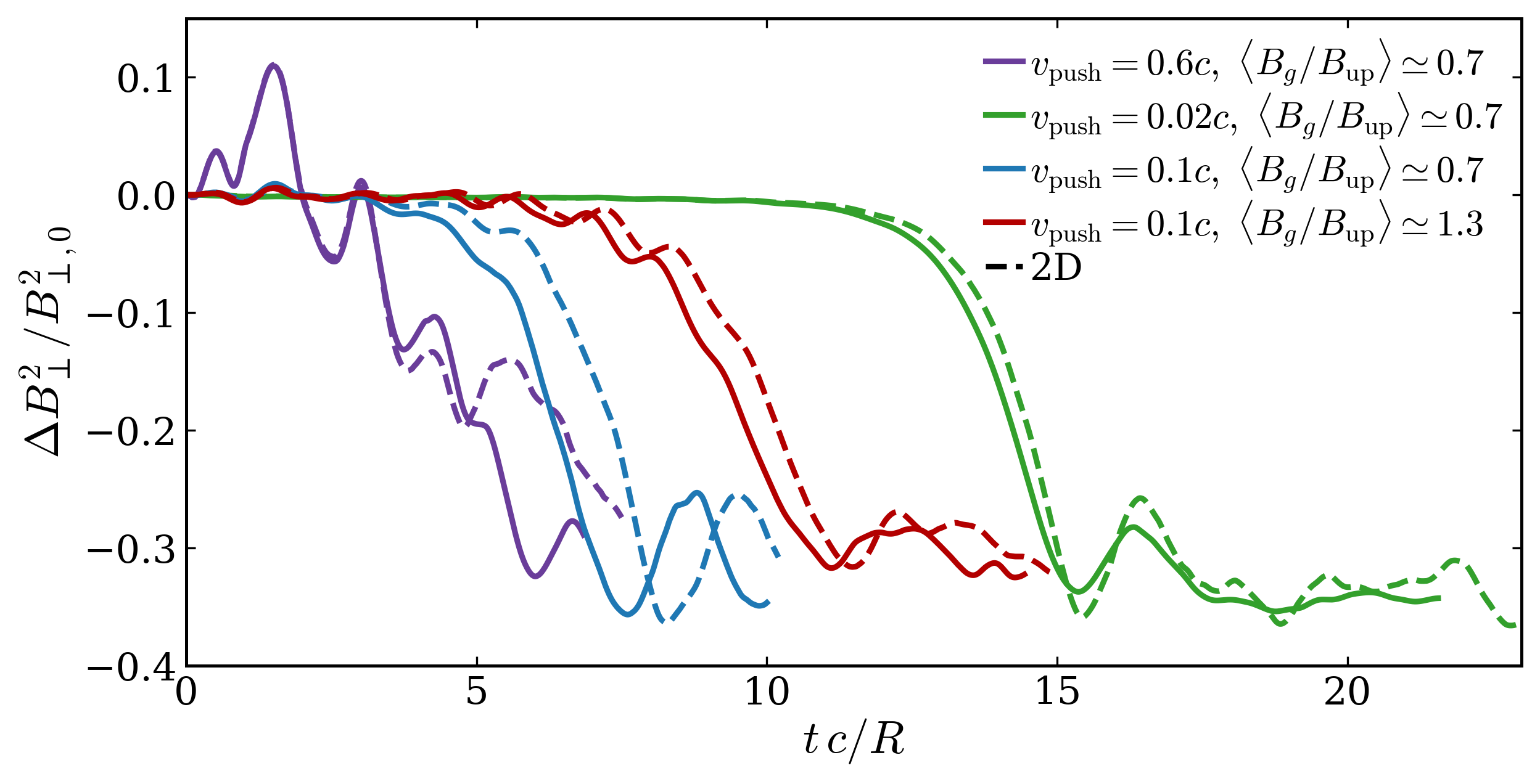}
    \caption{Temporal evolution of the perpendicular magnetic energy depletion, $\varepsilon_{B_\perp} = \big(B_\perp^2(t)-B_{\perp,0}^2\big)/B_{\perp,0}^2$, computed from the volume-integrated perpendicular magnetic energy over the entire simulation domain, for 3D (solid) and 2D (dashed).}
    \label{fig:energy}
\end{figure}

Despite the markedly different early-time behavior and characteristic timescales, Fig.~\ref{fig:cutoff} shows that the high-energy cutoff of accelerated particles converges to the same asymptotic value in all runs, with a maximum normalized cutoff $\left(\gamma_{\rm cut}/\sigma_{\rm in}\right)_{\rm max}\simeq 50$. This value exhibits only a very weak dependence on dimensionality (2D versus 3D), on the imposed drive strength $v_{\rm push}$, and on the guide-field amplitude within the parameter range explored here. To assess the role of the system size more explicitly, we include in Appendix~\ref{app:sys_size} an additional figure showing the evolution of $\gamma_{\rm cut}$ as a function of $R/d_e$.

This convergence can be understood from an electric-field-limited acceleration argument. While a particle remains in the energizing region, its Lorentz factor satisfies $\rmd\gamma/\rmd t=(e/m_ec^2)\,\boldsymbol{E}\!\bcdot\!\boldsymbol{v}\lesssim eE_{\rm rec}/(m_ec)$, so the maximum attainable energy gain is set by the reconnection electric field integrated over the effective acceleration time, $\gamma_{\rm cut}\sim (eE_{\rm rec}/m_ec)\,\Delta t_{\rm acc}$. Adopting the standard collisionless-reconnection scaling $E_{\rm rec}\simeq\epsilon\,(v_{A,\rm out}/c)\,B_{\rm up}$ with $\epsilon\sim0.1$ \citep[e.g.,][]{Guo2014,Sironi2014,Werner2016,Werner2017}, the late-time cutoff in our setup depends primarily on the in-plane magnetization $\sigma_{\rm in}$ and on the duration of the fast-merging stage $\Delta t_{\rm acc}$. Since $R/d_e$ and $\sigma_0$ (hence $\sigma_{\rm in}$) are fixed across our runs, and $\Delta t_{\rm acc}c/R\simeq2$--$3$ varies only weakly with $v_{\rm push}$, the normalized cutoff naturally converges to a constant within our parameter scan. Using representative values for our simulations yields $\gamma_{\rm cut}/\sigma_{\rm in}\sim30$--$60$ for $\epsilon\sim0.1$, in quantitative agreement with the measured asymptote.

The 3D runs accelerate particles slightly earlier than the 2D runs (Fig.~\ref{fig:cutoff}). As shown in Fig.~\ref{fig:energy}, the perpendicular magnetic energy is depleted more rapidly in 3D during the merging phase, providing earlier access to the reconnection electric field that drives particle energization.

The acceleration rate depends both on the driving strength and on the evolutionary stage. We distinguish two phases: an early stage (Phases I–II), during which the current sheet forms and reconnection proceeds slowly, and the fast-merging stage (Phase III), during which most of the energization occurs.
For strong driving ($v_{\rm push}=0.6c$), the early stage is characterized by rapid fractional growth, $\langle \rmd\ln\gamma_{\rm cut}/\rmd t\rangle \simeq 2.9\,(c/R)$, which drops to $\simeq 0.3\,(c/R)$ during the merging phase. For weak driving ($v_{\rm push}=0.02c$), the early-stage growth is slower, $\simeq 0.45\,(c/R)$, while the merging phase dominates the energization, with $\langle \rmd\ln\gamma_{\rm cut}/\rmd t\rangle \simeq 0.58\,(c/R)$ and $\rmd\gamma_{\rm cut}/\rmd t \simeq 12.1$.
During the fast-merging phase, $\gamma_{\rm cut}$ grows approximately linearly in time, $\gamma_{\rm cut}\simeq a\,(ct/R)+b$, with $a\simeq 13.2,\,10.8,\,9.8$ for $v_{\rm push}=0.02c,\,0.1c,\,0.6c$, respectively. This linear behavior indicates a nearly constant acceleration rate set by the reconnection electric field, whereas the earlier phase reflects transient, drive-dependent dynamics.

\begin{table}[h!]
\centering
\caption{Fitted power-law indices of the particle spectra. The index $p$ is defined from ${\rm d}N/{\rm d}\gamma \propto \gamma^{-p}$, with fits performed over $3 \leq \gamma \leq 70$.}
\label{tab:spectral_indices}
\begin{tabular}{lc}
\hline
Case & Index $p$ \\
\hline
$\langle B_g/B_{\rm up}\rangle\simeq 1.3,\; v_{\rm push}=0.1c,\; 3\mathrm{D}$        & $1.70$ \\
$\langle B_g/B_{\rm up}\rangle\simeq 1.3,\; v_{\rm push}=0.1c,\; 2\mathrm{D}$        & $1.82$ \\
$\langle B_g/B_{\rm up}\rangle\simeq 0.7,\; v_{\rm push}=0.1c,\; 3\mathrm{D}$        & $1.71$ \\
$\langle B_g/B_{\rm up}\rangle\simeq 0.7,\; v_{\rm push}=0.1c,\; 2\mathrm{D}$        & $1.86$ \\
$\langle B_g/B_{\rm up}\rangle\simeq 0.7,\; v_{\rm push}=0.6c,\; 3\mathrm{D}$        & $1.87$ \\
$\langle B_g/B_{\rm up}\rangle\simeq 0.7,\; v_{\rm push}=0.6c,\; 2\mathrm{D}$        & $2.00$ \\
$\langle B_g/B_{\rm up}\rangle\simeq 0.7,\; v_{\rm push}=0.02c,\; 3\mathrm{D}$       & $1.58$ \\
$\langle B_g/B_{\rm up}\rangle\simeq 0.7,\; v_{\rm push}=0.02c,\; 2\mathrm{D}$       & $1.76$ \\
\hline
\end{tabular}
\end{table}

\begin{figure}
    \centering
    \includegraphics[width=1\linewidth]{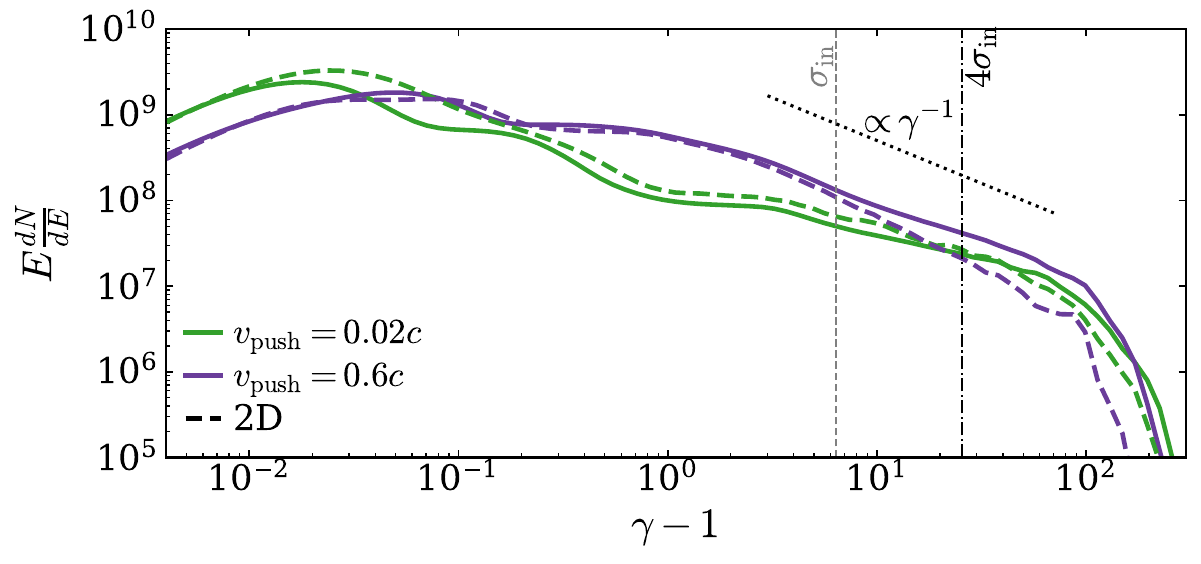}
    \caption{Spectra for the low–guide-field runs ($\langle B_g/B_{\rm up}\rangle\simeq0.7$) with $v_{\rm push}=0.02c$ (green) and $0.6c$ (purple), comparing 3D (solid) and 2D (dashed) simulations. Vertical lines mark $\sigma_{\rm in}$ and $4\sigma_{\rm in}$, and the dotted line indicates a $\gamma^{-1}$ reference slope.}
    \label{fig:spectra_p}
\end{figure}

The particle energy spectra for runs $v_{\rm push}=0.02c$ and $v_{\rm push}=0.6c$ are shown in Fig.~\ref{fig:spectra_p}. Power-law indices $p$, measured over the nonthermal range indicated in the figure, are listed in Table~\ref{tab:spectral_indices}. Across all simulations, the indices vary only modestly, with $p\simeq1.6$--$2.0$.

\subsection{Acceleration in the Reconnection Layer}

\begin{figure}
    \centering
    \includegraphics[width=1\linewidth]{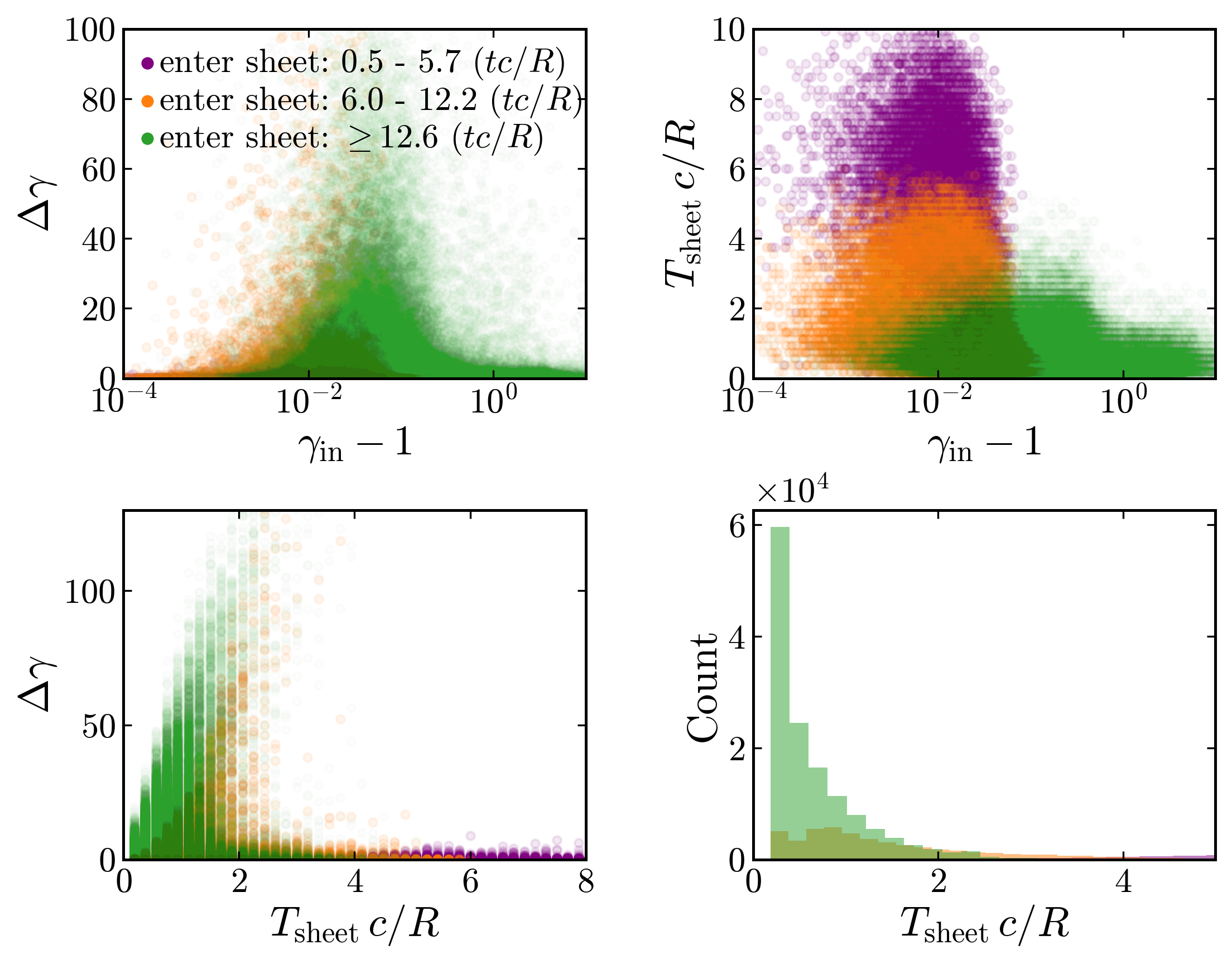}
    \caption{Particle energization statistics for the 3D run with $v_{\rm push}=0.02c$ and $\langle B_g/B_{\rm up}\rangle=0.7$. As described in the text, particles are grouped according to the three dynamical stages: entries during Phase~I (purple points), during Phase~II (orange points), and during Phase~III (green points). 
    Top left: energy gain $\Delta\gamma=\gamma_{\rm out}-\gamma_{\rm in}$ as a function of injection energy $\gamma_{\rm in}-1$. 
    Top right: residence time in the current sheet $T_{\rm sheet}$ versus $\gamma_{\rm in}-1$. 
    Bottom left: $\Delta\gamma$ versus $T_{\rm sheet}$. 
    Bottom right: distributions of $T_{\rm sheet}$. }
    \label{fig:stats_vp002}
\end{figure}

\begin{figure}
    \centering
    \includegraphics[width=1\linewidth]{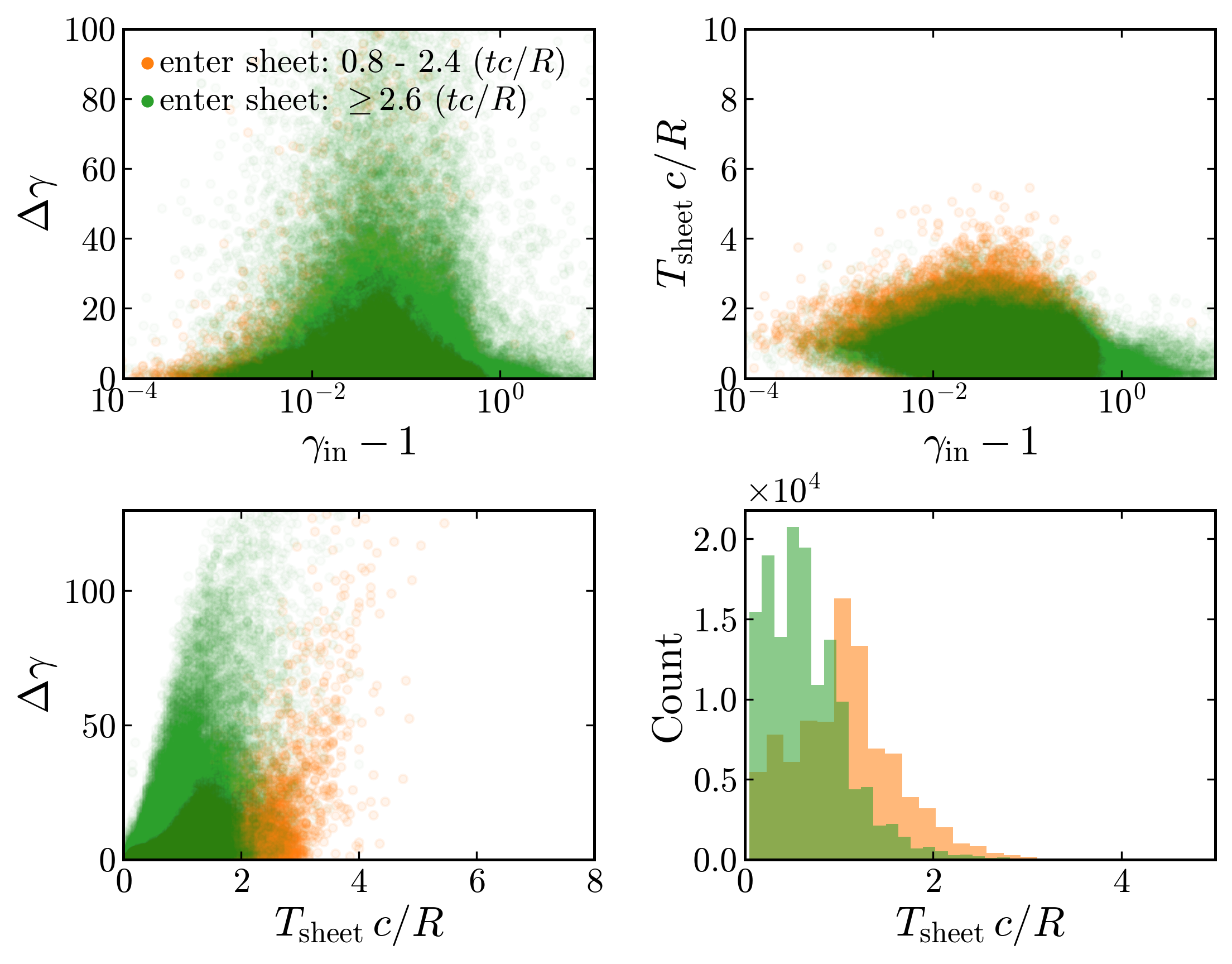}
    \caption{Same diagnostics as Fig.~\ref{fig:stats_vp002}, but for the 3D run with $v_{\rm push}=0.6c$ and $\langle B_g/B_{\rm up}\rangle=0.7$. Here, particle entry times are split into two stages.}
    \label{fig:stats_vp06}
\end{figure}

We track individual particles and, for each continuous residence interval inside the current sheet, record the Lorentz factor at entry and exit ($\gamma_{\rm in}$, $\gamma_{\rm out}$), the net gain $\Delta\gamma\equiv\gamma_{\rm out}-\gamma_{\rm in}$, and the residence time $T_{\rm sheet}$ (in units of $R/c$; computed in both 2D and 3D). We then condition these statistics on the entry time $t_{\rm in}$ to relate particle injection to the global evolution. For the slow-drive case ($v_{\rm push}=0.02c$; Fig.~\ref{fig:stats_vp002}), particles are separated into three stages: Phase~I (sheet formation; purple), Phase~II (drive-dominated), and Phase~III (merging; green). For the fast-drive case ($v_{\rm push}=0.6c$; Fig.~\ref{fig:stats_vp06}), the distribution splits into two groups (Phase II starts almost immediately): early entries during the drive-dominated stage (orange) and later entries once reconnection and merging dominate (green). In both figures $\langle B_g/B_{\rm up}\rangle=0.7$, and the four panels show (top left) $\Delta\gamma$ vs.\ $\gamma_{\rm in}-1$, (top right) $T_{\rm sheet}$ vs.\ $\gamma_{\rm in}-1$, (bottom left) $\Delta\gamma$ vs.\ $T_{\rm sheet}$, and (bottom right) the distribution of $T_{\rm sheet}$.
 
In 3D runs, a very large number of particles cross the sheet at least once ($\sim 2.1 \times 10^5$ for $v_{\rm push}=0.02c$ and $\sim 2.7 \times 10^5$ for $v_{\rm push}=0.6c$). These correspond to \(\sim 10.3\%\) and \(\sim 11.2\%\) of the full tracked particle population. Typical entry energies are only mildly relativistic, $\langle\gamma_{\rm in}\rangle\simeq1.1$--$1.2\ll\sigma_{\rm in}$, while exit energies remain well below the global magnetization scale: $\langle\gamma_{\rm out}\rangle/\sigma_{\rm in}\simeq0.88$ at $v_{\rm push}=0.6c$ and $\simeq0.78$ at $0.02c$. The corresponding mean energy gains are $\langle\Delta\gamma\rangle/\sigma_{\rm in}\simeq0.70$ and $\simeq0.59$, respectively. 

A key result is that, in our simulations, the reconnection layer acts as a largely injection-energy-agnostic accelerator. Across all runs (3D and 2D, slow and fast push), the correlation between $\Delta\gamma$ and $\gamma_{\rm in}$ remains weak. The resulting $\Delta\gamma$–$\gamma_{\rm in}$ distributions form broad, roughly Gaussian clouds whose centroid and width show only a shallow dependence on $\gamma_{\rm in}-1$. Notably, the late-time populations in the fast-merging regime (green points) behave almost identically for $v_{\rm push}=0.02c$ and $0.6c$, yielding the same characteristic $\Delta\gamma$ and comparable spread. Particles that enter the sheet with higher initial energies tend, if anything, to experience smaller fractional energization than initially cold particles.

The $\Delta\gamma$--$T_{\rm sheet}$ panel (bottom left) shows a sharp upper envelope, with essentially no particles exceeding a quasilinear cutoff, i.e., a maximum energy set by direct acceleration in the reconnection electric field over the available residence time. This envelope reflects a hard limit on the acceleration rate imposed by the reconnection electric field acting in the X-point region. In this injection-dominated regime, particle energization is controlled by direct acceleration by $E_{\parallel}$, with a rate bounded by
\[
\frac{\rmd\gamma}{\rmd t}\lesssim \frac{eE_{\rm rec}R}{m_ec^2}.
\]
This bound naturally produces the sharp upper envelope $\Delta\gamma\propto T_{\rm sheet}$.
Using $E_{\rm rec}=\epsilon(v_{A,\rm out}/c)B_{\rm up}$ with $\epsilon\simeq0.1$, we obtain a dimensionless acceleration rate $\alpha_0=\epsilon(v_{A,\rm out}/c)\sqrt{\sigma_{\rm up}}(R/d_e)$, in quantitative agreement with the measured slopes in both 2D and 3D.

The $\vp = 0.02c$  3D run shows a much broader early population in both time and energy. The orange points in the $T_{\rm sheet}$–$(\gamma_{\rm in}-1)$ plane form a high-density cloud extending to $T_{\rm sheet}\sim 10 R/c$ for injection energies $10^{-3}\lesssim\gamma_{\rm in}-1\lesssim 10^{-1}$. 

The 2D runs (not shown) follow the same qualitative trends but with systematically smaller gains. For $v_{\rm push}=0.6c$ we find $\langle\Delta\gamma\rangle/\sigma_{\rm in}\simeq0.31$, $\langle T_{\rm sheet}\rangle\simeq0.96 R/c$, while for $v_{\rm push}=0.02c$ the mean gain increases to $\langle\Delta\gamma\rangle/\sigma_{\rm in}\simeq0.42$.

A stronger guide field lowers the injection energy of particles entering the sheet: we measure $\langle\gamma_{\rm in}\rangle \simeq 1.04$, compared to $1.14$--$1.22$ in weak-guide-field runs. In these runs, the guide-field ratio reaches $\langle B_g/B_{\rm up}\rangle \simeq 1.3$ during the transient phase. The stronger guide field also reduces the average energization per crossing by $\sim$10--30\% and shortens particle residence times in the sheet, while strongly suppressing repeated entries and long trapping episodes.

Overall, these statistics show that, in our simulations, (i) acceleration in the main current layer is controlled primarily by the residence time and internal field structure, consistent with direct energization by the reconnection electric field rather than by the injection energy; (ii) weak driving produces a long-lived premerger current sheet (Phase II) in which a small fraction of particles remains trapped for extended times and undergoes substantial acceleration; and (iii) 3D geometry enhances this effect slightly compared to 2D by allowing longer residence times and, in weak-guide-field cases, more repeated interactions with the accelerating regions.

\section{Conclusion}
\label{sec:conclusion}

We presented 3D PIC simulations of driven magnetic reconnection produced self-consistently by the compression and merging of two force-free flux tubes in a strongly magnetized electron–positron plasma with relativistic Alfv\'enic outflows. The early evolution remains quasi-2D in a $z$-averaged sense during sheet formation, but reconnection onset is systematically delayed in 3D relative to corresponding 2D runs. Increasing the guide field also delays onset. In the weak-guide-field case ($\langle B_g/B_{\rm up}\rangle\simeq0.7$), spectra in $(k_x,k_z)$ show oblique power during thinning; for stronger guide field ($\langle B_g/B_{\rm up}\rangle\simeq1.3$), oblique components are suppressed and the dynamics becomes more anisotropic and effectively 2C.

The imposed drive controls linear growth in both instability channels. For tearing, the measured growth rates increase almost linearly with $v_{\rm push}$: $\gamma_{\rm tear}\simeq0.61,2.0,4.19c/R$ for $v_{\rm push}/c=0.02,0.1,0.6$ (weak guide field), with similar saturation amplitudes. Drift-kink grows faster when it is present ($\gamma_{\rm kink}\simeq7.12,6.7,13.43c/R$ for the same three pushes) but it appears after the first signs of reconnection, rather than setting the onset. The guide field primarily regulates the drift-kink instability: at $v_{\rm push}=0.1c$, increasing $\langle B_g/B_{\rm up}\rangle$ from $\simeq0.7$ to $\simeq1.3$ removes coherent kink growth, while the tearing growth rate is only mildly reduced. During fast merging, the normalized reconnection rate reaches $\sim0.08$ on average with peaks up to $\sim0.1$.

Particle acceleration is robust across the parameter scan. The high-energy cutoff converges to the same asymptotic value in all runs, $(\gamma_{\rm cut}/\sigma_{\rm in})_{\max}\simeq50$, with weak dependence on 2D vs.\ 3D, $v_{\rm push}$, and guide-field strength in the explored range. The nonthermal spectra remain similar across runs, with fitted indices defined from ${\rm d}N/{\rm d}\gamma \propto \gamma^{-p}$ over $3\le\gamma\le70$, spanning $p\sim 1.6$--$3.0$ (e.g., at $v_{\rm push}=0.1c$: $p\simeq1.71$ in 3D vs.\ $1.86$ in 2D for weak guide; $p\simeq1.71$ in 3D vs.\ $1.83$ in 2D for strong guide field). 3D runs produce slightly earlier and slightly faster energization, but do not not change the late-time cutoff or the overall spectral shape compared to 2D for this driven flux-tube geometry.

\begin{acknowledgments}
C.G.~thanks D. Uzdensky, L. Comisso, E. Gorbunov, for their helpful comments.
F.B.\ acknowledges support from the FED-tWIN programme (profile Prf-2020-004, project ``ENERGY''), issued by BELSPO, and from the FWO Junior Research Project G020224N granted by the Research Foundation -- Flanders (FWO).
D.G.~is supported by the Research Foundation--Flanders (FWO) Senior Postdoctoral Fellowship 12B1424N.
The resources and services used in this work were provided by the VSC (Flemish Supercomputer Center),  funded by the FWO and the Flemish Government.
L.S. acknowledges support from DoE Early Career Award DE-SC0023015, NASA ATP 80NSSC24K1238, NASA ATP 80NSSC24K1826, and NSF AST-2307202. This work was supported by a grant from the Simons Foundation (MP-SCMPS-00001470) to L.S. and facilitated by Multimessenger Plasma Physics Center (MPPC), grant NSF PHY-2206609 to L.S.
\end{acknowledgments}
{\software{\textsc{Tristan-MP v2} \citep{tristan}}}

\appendix

\section{Control Parameter $C$ and Force-Free Equilibrium}
\label{app:cparam}

The dimensionless control parameter $C$ sets the relative strength of the axial component $B_z$ with respect to the reconnecting (poloidal) component $B_\phi$. For $C=0$, the configuration is exactly force-free ($\bb{J}\btimes\bb{B}=\bb{0}$), while finite $C$ introduces a small deviation from force-free balance by modifying the proportionality between the magnetic field and its curl.

For $C\neq0$, a weak outward radial Lorentz force appears,
$\bb{f}(r)=\bb{J}\btimes\bb{B}$,
leading to a brief initial readjustment through fast-magnetosonic and Alfvén waves on a timescale of the order of the Alfvén crossing time, $\tau_A\sim R/c$. In practice, the value used in this work ($C=0.2$) corresponds only to a small departure from force-free equilibrium and does not qualitatively alter the initial flux-tube structure or subsequent reconnection dynamics. We have verified this over a wider range of values, testing configurations up to $C=2$.

\section{Impact of Driving on Fastest Growth Rate}
\label{app:fastest_modes}

\begin{figure}[t]
    \centering
    \includegraphics[width=0.48\linewidth]{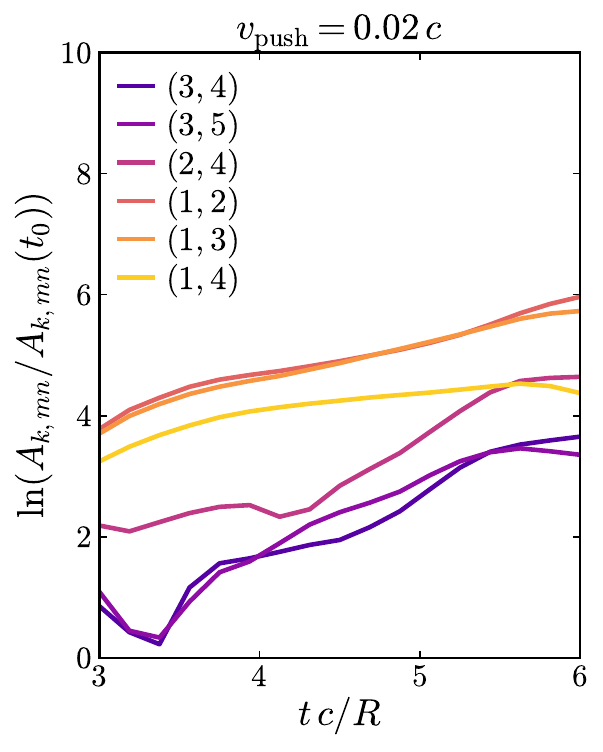}
    \hfill
    \includegraphics[width=0.48\linewidth]{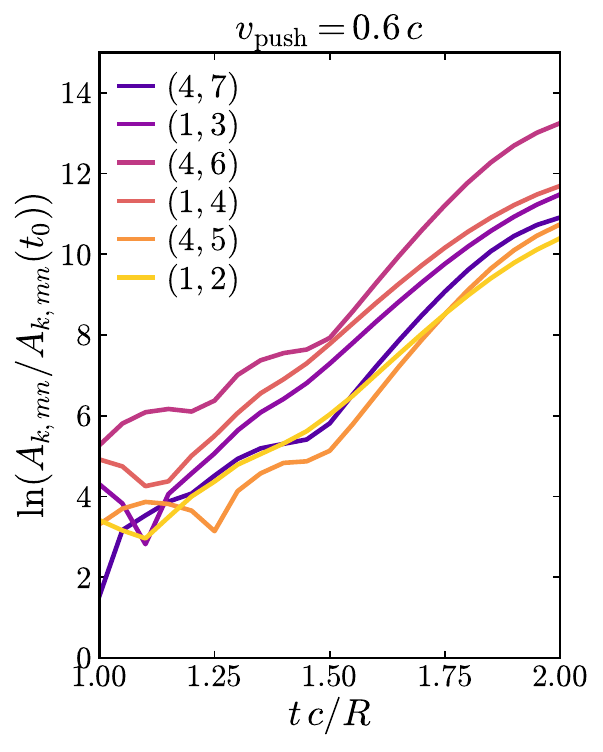}
    \caption{Time evolution of the six fastest-growing Fourier mode pairs $(m_x,m_z)$ for slow ($v_{\rm push}=0.02c$, left) and fast ($v_{\rm push}=0.6c$, right) driving during their linear phase.}
    \label{fig:fastest}
\end{figure}

In this appendix, we provide a complementary view of how the external compression affects the growth of individual Fourier modes. Fig.~\ref{fig:fastest} shows the time evolution of the six fastest-growing mode pairs $(m_x,m_z)$ for a slow drive ($v_{\rm push}=0.02c$) and a fast drive ($v_{\rm push}=0.6c$), over time intervals where the amplitudes display approximately exponential growth.

For the slow drive, the fastest modes span a wide range of growth rates ($\gamma \simeq 0.3$--$1.1\,c/R$) and wavenumbers, with no single mode clearly separated from the rest. In contrast, for the fast drive all dominant modes grow rapidly with tightly clustered rates ($\gamma \simeq 7.4$--$8.9\,c/R$), indicating that increasing the push uniformly accelerates the growth of a broad spectrum of modes rather than selectively amplifying a specific $(m_x,m_z)$ pair.

\vspace{0.5cm}

\section{System-size Dependence of the High-energy Cutoff}
\label{app:sys_size}

To assess the system-size dependence of the high-energy extent, we compare 2D simulations with identical plasma parameters and different values of $R/d_e$. At fixed $\sigma_{\rm in}=6.4$, the maximum Lorentz factor inferred from the highest occupied spectral bin increases monotonically with system size, from $\gamma_{\max}/\sigma_{\rm in}\simeq 40.4$ at $R/d_e\simeq 138.7$ to $\simeq 79.9$ at $R/d_e\simeq 533.3$. Expressed in terms of the kinetic size parameter $R/r_{L,\rm hot}$, with $r_{L,\rm hot}=\sqrt{\sigma_{\rm in}}\,d_e$, the equal-time trend is well described by a sublinear scaling, $\gamma_{\max}/\sigma_{\rm in}\propto (R/r_{L,\rm hot})^{0.5}$. The early-time evolution is nearly independent of system size, while the differences emerge primarily during the late fast-merging phase, where larger systems continue to extend the spectrum to higher energies. 

\vspace{0.5cm}
\begin{figure}
    \centering
    \includegraphics[width=1\linewidth]{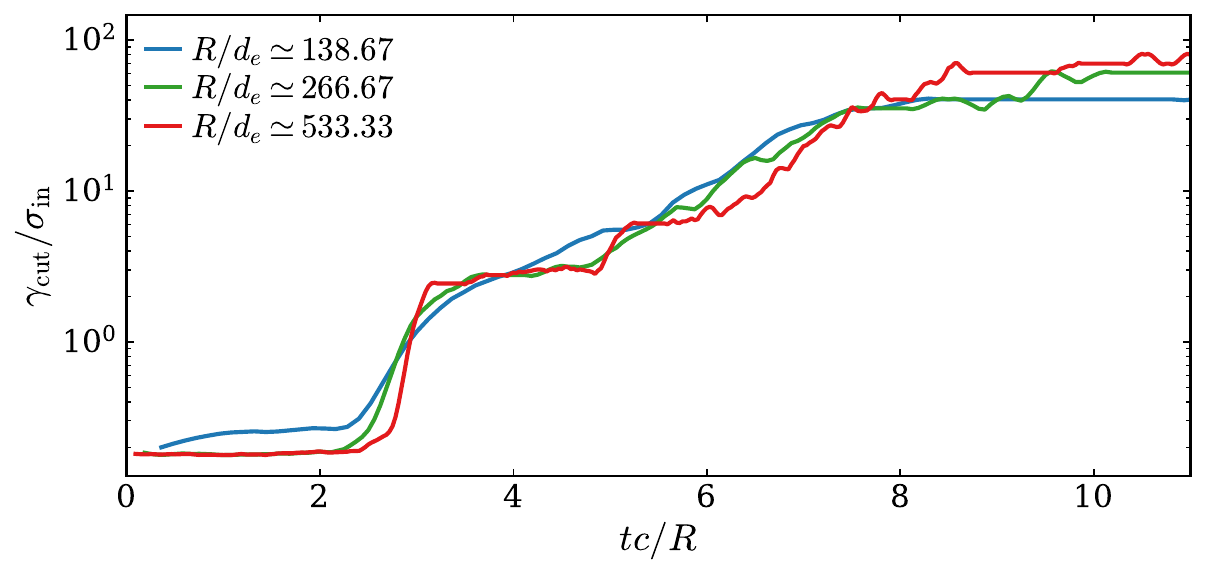}
    \caption{Temporal evolution of the maximum particle Lorentz factor, $\gamma_{\rm cut}$, normalized to the injected magnetization $\sigma_{\rm in}$, for four 2D simulations with different system sizes. }
    \label{fig:gamma_max_size}
\end{figure}

\bibliography{biblio-3}

@ARTICLE{Cerutti2013,
       author = {{Cerutti}, B. and {Werner}, G.~R. and {Uzdensky}, D.~A. and {Begelman}, M.~C.},
        title = "{Simulations of Particle Acceleration beyond the Classical Synchrotron Burnoff Limit in Magnetic Reconnection: An Explanation of the Crab Flares}",
      journal = {\apj},
         year = 2013,
        month = jun,
       volume = {770},
       number = {2},
          eid = {147},
        pages = {147},
          doi = {10.1088/0004-637X/770/2/147},
       adsurl = {https://ui.adsabs.harvard.edu/abs/2013ApJ...770..147C}
}

@article{Uzdensky2011,
doi = {10.1088/2041-8205/737/2/L40},
url = {https://dx.doi.org/10.1088/2041-8205/737/2/L40},
year = {2011},
month = {aug},
publisher = {The American Astronomical Society},
volume = {737},
number = {2},
pages = {L40},
author = {Uzdensky, Dmitri A. and Cerutti, Benoît and Begelman, Mitchell C.},
title = {RECONNECTION-POWERED LINEAR ACCELERATOR AND GAMMA-RAY FLARES IN THE CRAB NEBULA},
journal = {The Astrophysical Journal Letters},
}

@ARTICLE{Loureiro2007,
       author = {{Loureiro}, N.~F. and {Schekochihin}, A.~A. and {Cowley}, S.~C.},
        title = "{Instability of current sheets and formation of plasmoid chains}",
      journal = {Physics of Plasmas},
         year = 2007,
        month = oct,
       volume = {14},
       number = {10},
        pages = {100703-100703},
          doi = {10.1063/1.2783986},
archivePrefix = {arXiv},
       eprint = {astro-ph/0703631},
 primaryClass = {astro-ph},
       adsurl = {https://ui.adsabs.harvard.edu/abs/2007PhPl...14j0703L}
}

@article{Lapenta2008,
  title = {Self-Feeding Turbulent Magnetic Reconnection on Macroscopic Scales},
  author = {Lapenta, Giovanni},
  journal = {Phys. Rev. Lett.},
  volume = {100},
  issue = {23},
  pages = {235001},
  numpages = {4},
  year = {2008},
  month = {Jun},
  publisher = {American Physical Society},
  doi = {10.1103/PhysRevLett.100.235001},
  url = {https://link.aps.org/doi/10.1103/PhysRevLett.100.235001}
}

@article{Lyutikov2017, title={Particle acceleration in relativistic magnetic flux-merging events}, volume={83}, DOI={10.1017/S002237781700071X}, number={6}, journal={Journal of Plasma Physics}, author={Lyutikov, Maxim and Sironi, Lorenzo and Komissarov, Serguei S. and Porth, Oliver}, year={2017}, pages={635830602}}

@ARTICLE{Ripperda2017a,
       author = {{Ripperda}, B. and {Porth}, O. and {Xia}, C. and {Keppens}, R.},
        title = "{Reconnection and particle acceleration in interacting flux ropes - I. Magnetohydrodynamics and test particles in 2.5D}",
      journal = {\mnras},
         year = 2017,
        month = may,
       volume = {467},
       number = {3},
        pages = {3279-3298},
          doi = {10.1093/mnras/stx379},
archivePrefix = {arXiv},
       eprint = {1611.09966},
 primaryClass = {astro-ph.HE},
       adsurl = {https://ui.adsabs.harvard.edu/abs/2017MNRAS.467.3279R}
}

@article{Huang2017,
doi = {10.3847/1538-4357/aa906d},
url = {https://dx.doi.org/10.3847/1538-4357/aa906d},
year = {2017},
month = {nov},
publisher = {The American Astronomical Society},
volume = {849},
number = {2},
pages = {75},
author = {Huang, Yi-Min and Comisso, Luca and Bhattacharjee, A.},
title = {Plasmoid Instability in Evolving Current Sheets and Onset of Fast Reconnection},
journal = {The Astrophysical Journal}
}

@ARTICLE{Sironi2014,
       author = {{Sironi}, Lorenzo and {Spitkovsky}, Anatoly},
        title = "{Relativistic Reconnection: An Efficient Source of Non-thermal Particles}",
      journal = {\apjl},
         year = 2014,
        month = mar,
       volume = {783},
       number = {1},
          eid = {L21},
        pages = {L21},
          doi = {10.1088/2041-8205/783/1/L21},
       adsurl = {https://ui.adsabs.harvard.edu/abs/2014ApJ...783L..21S}
}

@ARTICLE{Comisso2016,
       author = {{Comisso}, L. and {Bhattacharjee}, A.},
        title = "{On the value of the reconnection rate}",
      journal = {Journal of Plasma Physics},
         year = 2016,
        month = dec,
       volume = {82},
       number = {6},
          eid = {595820601},
        pages = {595820601},
          doi = {10.1017/S002237781600101X},
       adsurl = {https://ui.adsabs.harvard.edu/abs/2016JPlPh..82f5901C}
}

@ARTICLE{Werner2016,
       author = {{Werner}, G.~R. and {Uzdensky}, D.~A. and {Cerutti}, B. and {Nalewajko}, K. and {Begelman}, M.~C.},
        title = "{The Extent of Power-law Energy Spectra in Collisionless Relativistic Magnetic Reconnection in Pair Plasmas}",
      journal = {\apjl},
         year = 2016,
        month = jan,
       volume = {816},
       number = {1},
          eid = {L8},
        pages = {L8},
          doi = {10.3847/2041-8205/816/1/L8},
       adsurl = {https://ui.adsabs.harvard.edu/abs/2016ApJ...816L...8W}
}

@article{Granier2022,
  title = {Marginally stable current sheets in collisionless magnetic reconnection},
  author = {Granier, C. and Borgogno, D. and Comisso, L. and Grasso, D. and Tassi, E. and Numata, R.},
  journal = {Phys. Rev. E},
  volume = {106},
  issue = {4},
  pages = {L043201},
  numpages = {5},
  year = {2022},
  month = {Oct},
  publisher = {American Physical Society},
  doi = {10.1103/PhysRevE.106.L043201},
  url = {https://link.aps.org/doi/10.1103/PhysRevE.106.L043201}
}

@article{Hakobyan2023,
doi = {10.3847/1538-4357/acab05},
url = {https://dx.doi.org/10.3847/1538-4357/acab05},
year = {2023},
month = {jan},
publisher = {The American Astronomical Society},
volume = {943},
number = {2},
pages = {105},
author = {{Hakobyan}, Hayk and Philippov, Alexander and Spitkovsky, Anatoly},
title = {Magnetic Energy Dissipation and γ-Ray Emission in Energetic Pulsars},
journal = {The Astrophysical Journal}
}

@ARTICLE{Sironi2025,
       author = {{Sironi}, Lorenzo and {Uzdensky}, Dmitri A. and Giannios, Dimitrios},
        title = "{Relativistic Magnetic Reconnection in
Astrophysical Plasmas: A Powerful Mechanism of Nonthermal Emission}",
      journal = {\araa},
         year = 2025,
        month = may,
       volume = {63},
          doi = {10.1146/annurev-astro-020325-115713},
       adsurl = {https://ui.adsabs.harvard.edu/abs/2006ARA&A..44...49R},
}

@ARTICLE{Hakobyan2023b,
       author = {{Hakobyan}, Hayk and {Ripperda}, B. and {Philippov}, A.~A.},
        title = "{Radiative Reconnection-powered TeV Flares from the Black Hole Magnetosphere in M87}",
      journal = {\apjl},
         year = 2023,
        month = feb,
       volume = {943},
       number = {2},
          eid = {L29},
        pages = {L29},
          doi = {10.3847/2041-8213/acb264},
       adsurl = {https://ui.adsabs.harvard.edu/abs/2023ApJ...943L..29H}
}

@MISC{tristan,
       author = {{Hakobyan}, Hayk and {Spitkovsky}, Anatoly and {Chernoglazov}, Alexander and {Philippov}, Alexander and {Groselj}, Daniel and {Mahlmann}, Jens},
        title = "{PrincetonUniversity/tristan-mp-v2: v2.6}",
         year = 2023,
        month = jan,
          eid = {10.5281/zenodo.7566725},
          doi = {10.5281/zenodo.7566725},
    publisher = {Zenodo},
       adsurl = {https://ui.adsabs.harvard.edu/abs/2023zndo...7566725H}
}

@article{Hakobyan2021,
doi = {10.3847/1538-4357/abedac},
url = {https://doi.org/10.3847/1538-4357/abedac},
year = {2021},
month = {may},
publisher = {The American Astronomical Society},
volume = {912},
number = {1},
pages = {48},
author = {Hakobyan, Hayk and Petropoulou, Maria and Spitkovsky, Anatoly and Sironi, Lorenzo},
title = {Secondary Energization in Compressing Plasmoids during Magnetic Reconnection},
journal = {The Astrophysical Journal}
}

@article{Barkov2016,
    author = {Barkov, Maxim V. and Komissarov, Serguei S.},
    title = {Relativistic tearing and drift-kink instabilities in two-fluid simulations},
    journal = {Monthly Notices of the Royal Astronomical Society},
    volume = {458},
    number = {2},
    pages = {1939-1947},
    year = {2016},
    month = {03},
    issn = {0035-8711},
    doi = {10.1093/mnras/stw384},
    url = {https://doi.org/10.1093/mnras/stw384},
    eprint = {https://academic.oup.com/mnras/article-pdf/458/2/1939/18240261/stw384.pdf},
}

@article{Zhang2021,
  title = {Efficient Nonthermal Ion and Electron Acceleration Enabled by the Flux-Rope Kink Instability in 3D Nonrelativistic Magnetic Reconnection},
  author = {Zhang, Qile and Guo, Fan and Daughton, William and Li, Hui and Li, Xiaocan},
  journal = {Phys. Rev. Lett.},
  volume = {127},
  issue = {18},
  pages = {185101},
  numpages = {7},
  year = {2021},
  month = {Oct},
  publisher = {American Physical Society},
  doi = {10.1103/PhysRevLett.127.185101},
  url = {https://link.aps.org/doi/10.1103/PhysRevLett.127.185101}
}

@article{Werner2017,
doi = {10.3847/2041-8213/aa7892},
url = {https://doi.org/10.3847/2041-8213/aa7892},
year = {2017},
month = {jul},
publisher = {The American Astronomical Society},
volume = {843},
number = {2},
pages = {L27},
author = {Werner, Gregory R. and Uzdensky, Dmitri A.},
title = {Nonthermal Particle Acceleration in 3D Relativistic Magnetic Reconnection in Pair Plasma},
journal = {The Astrophysical Journal Letters}
}

@article{Kuznetsova1985,
doi = {10.1088/0741-3335/27/4/001},
url = {https://doi.org/10.1088/0741-3335/27/4/001},
year = {1985},
month = {apr},
publisher = {},
volume = {27},
number = {4},
pages = {363},
author = {M M Kuznetsova and L M Zeleny},
title = {Stability and structure of the perturbations of the magnetic surfaces in the magnetic transitional layers},
journal = {Plasma Physics and Controlled Fusion}
}

@article{Granier2025,
doi = {10.3847/1538-4357/ae0738},
url = {https://doi.org/10.3847/1538-4357/ae0738},
year = {2025},
month = {oct},
publisher = {The American Astronomical Society},
volume = {992},
number = {2},
pages = {193},
author = {Granier, Camille and Grošelj, Daniel and Comisso, Luca and Bacchini, Fabio},
title = {Driven Collisionless Reconnection of Force-free Flux Tubes: From Onset to Coalescence},
journal = {The Astrophysical Journal}
}

@article{ZweibelYamada2009,
   author = "Zweibel, Ellen G. and Yamada, Masaaki",
   title = "Magnetic Reconnection in Astrophysical and Laboratory Plasmas", 
   journal= "Annual Review of Astronomy and Astrophysics",
   year = "2009",
   volume = "47",
   number = "Volume 47, 2009",
   pages = "291-332",
   doi = "https://doi.org/10.1146/annurev-astro-082708-101726",
   url = "https://www.annualreviews.org/content/journals/10.1146/annurev-astro-082708-101726",
   publisher = "Annual Reviews",
   issn = "1545-4282",
   type = "Journal Article",
  }

@article{YamadaKulsrudJi2010,
  title = {Magnetic reconnection},
  author = {Yamada, Masaaki and Kulsrud, Russell and Ji, Hantao},
  journal = {Rev. Mod. Phys.},
  volume = {82},
  issue = {1},
  pages = {603--664},
  numpages = {0},
  year = {2010},
  month = {Mar},
  publisher = {American Physical Society},
  doi = {10.1103/RevModPhys.82.603},
  url = {https://link.aps.org/doi/10.1103/RevModPhys.82.603}
}

@article{JiDaughton2011,
    author = {Ji, Hantao and Daughton, William},
    title = {Phase diagram for magnetic reconnection in heliophysical, astrophysical, and laboratory plasmas},
    journal = {Physics of Plasmas},
    volume = {18},
    number = {11},
    pages = {111207},
    year = {2011},
    month = {10},
    issn = {1070-664X},
    doi = {10.1063/1.3647505},
    url = {https://doi.org/10.1063/1.3647505},
    eprint = {https://pubs.aip.org/aip/pop/article-pdf/doi/10.1063/1.3647505/14910995/111207_1_online.pdf},
}

@article{GuoEtAl2015,
title = {Particle acceleration and plasma dynamics during magnetic reconnection in the magnetically dominated regime},
author = {Guo, Fan and Liu, Yi -Hsin and Daughton, William and Li, Hui},
abstractNote = {Magnetic reconnection is thought to be the driver for many explosive phenomena in the universe. The energy release and particle acceleration during reconnection have been proposed as a mechanism for producing high-energy emissions and cosmic rays. We carry out two- and three-dimensional (3D) kinetic simulations to investigate relativistic magnetic reconnection and the associated particle acceleration. The simulations focus on electron–positron plasmas starting with a magnetically dominated, force-free current sheet (σ ≡ B2 / (4πnemec2) >> 1). For this limit, we demonstrate that relativistic reconnection is highly efficient at accelerating particles through a first-order Fermi process accomplished by the curvature drift of particles along the electric field induced by the relativistic flows. This mechanism gives rise to the formation of hard power-law spectra f α (γ - 1)-p and approaches p = 1 for sufficiently large σ and system size. Eventually most of the available magnetic free energy is converted into nonthermal particle kinetic energy. An analytic model is presented to explain the key results and predict a general condition for the formation of power-law distributions. The development of reconnection in these regimes leads to relativistic inflow and outflow speeds and enhanced reconnection rates relative to nonrelativistic regimes. In the 3D simulation, the interplay between secondary kink and tearing instabilities leads to strong magnetic turbulence, but does not significantly change the energy conversion, reconnection rate, or particle acceleration. This paper suggests that relativistic reconnection sites are strong sources of nonthermal particles, which may have important implications for a variety of high-energy astrophysical problems.},
doi = {10.1088/0004-637X/806/2/167},
journal = {The Astrophysical Journal (Online)},
number = 2,
volume = 806,
place = {United States},
year = {2015},
month = {6}
}

@article{WernerUzdensky2021, title={Reconnection and particle acceleration in three-dimensional current sheet evolution in moderately magnetized astrophysical pair plasma}, volume={87}, DOI={10.1017/S0022377821001185}, number={6}, journal={Journal of Plasma Physics}, author={Werner, Gregory R. and Uzdensky, Dmitri A.}, year={2021}, pages={905870613}}

@article{SironiBeloborodov2019,
doi = {10.3847/1538-4357/aba622},
url = {https://doi.org/10.3847/1538-4357/aba622},
year = {2020},
month = {aug},
publisher = {The American Astronomical Society},
volume = {899},
number = {1},
pages = {52},
author = {Sironi, Lorenzo and Beloborodov, Andrei M.},
title = {Kinetic Simulations of Radiative Magnetic Reconnection in the Coronae of Accreting Black Holes},
journal = {The Astrophysical Journal}
}

@article{Stanier2015,
    author = {Stanier, A. and Simakov, Andrei N. and Chacón, L. and Daughton, W.},
    title = {Fluid vs. kinetic magnetic reconnection with strong guide fields},
    journal = {Physics of Plasmas},
    volume = {22},
    number = {10},
    pages = {101203},
    year = {2015},
    month = {10},
    issn = {1070-664X},
    doi = {10.1063/1.4932330},
    url = {https://doi.org/10.1063/1.4932330},
    eprint = {https://pubs.aip.org/aip/pop/article-pdf/doi/10.1063/1.4932330/13526219/101203_1_online.pdf},
}

@article{Zenitani2008,
doi = {10.1086/528708},
url = {https://doi.org/10.1086/528708},
year = {2008},
month = {apr},
publisher = {},
volume = {677},
number = {1},
pages = {530},
author = {Zenitani, S. and Hoshino, M.},
title = {The Role of the Guide Field in Relativistic Pair Plasma Reconnection},
journal = {The Astrophysical Journal}
}

@article{Hoshino2024,
  author       = {Hoshino, Masahiro},
  title        = {A Hard Energy Spectrum in 3D Guide-Field Magnetic Reconnection},
  journal      = {arXiv e-prints},
  year         = {2024},
  eid          = {arXiv:2404.15662},
  eprint       = {2404.15662},
  primaryClass = {astro-ph.HE},
  doi          = {10.48550/arXiv.2404.15662},
  url   = {https://doi.org/10.48550/arXiv.2404.15662}
}

@article{Zenitani2005,
  title = {Three-Dimensional Evolution of a Relativistic Current Sheet: Triggering of Magnetic Reconnection by the Guide Field},
  author = {Zenitani, S. and Hoshino, M.},
  journal = {Phys. Rev. Lett.},
  volume = {95},
  issue = {9},
  pages = {095001},
  numpages = {4},
  year = {2005},
  month = {Aug},
  publisher = {American Physical Society},
  doi = {10.1103/PhysRevLett.95.095001},
  url = {https://link.aps.org/doi/10.1103/PhysRevLett.95.095001}
}

@article{Zenitani2007,
doi = {10.1086/522226},
url = {https://doi.org/10.1086/522226},
year = {2007},
month = {nov},
publisher = {},
volume = {670},
number = {1},
pages = {702},
author = {Zenitani, S. and Hoshino, M.},
title = {Particle Acceleration and Magnetic Dissipation in Relativistic Current Sheet of Pair Plasmas},
journal = {The Astrophysical Journal}
}

@ARTICLE{Guo2014,
       author = {{Guo}, Fan and {Li}, Hui and {Daughton}, William and {Liu}, Yi-Hsin},
        title = "{Formation of Hard Power Laws in the Energetic Particle Spectra Resulting from Relativistic Magnetic Reconnection}",
      journal = {\prl},
         year = 2014,
        month = oct,
       volume = {113},
       number = {15},
          eid = {155005},
        pages = {155005},
          doi = {10.1103/PhysRevLett.113.155005},
archivePrefix = {arXiv},
       eprint = {1405.4040},
 primaryClass = {astro-ph.HE},
       adsurl = {https://ui.adsabs.harvard.edu/abs/2014PhRvL.113o5005G}
}

@ARTICLE{Guo2015,
       author = {{Liu}, Yi-Hsin and {Guo}, Fan and {Daughton}, William and {Li}, Hui and {Hesse}, Michael},
        title = "{Scaling of Magnetic Reconnection in Relativistic Collisionless Pair Plasmas}",
      journal = {\prl},
         year = 2015,
        month = mar,
       volume = {114},
       number = {9},
          eid = {095002},
        pages = {095002},
          doi = {10.1103/PhysRevLett.114.095002},
archivePrefix = {arXiv},
       eprint = {1410.3178},
 primaryClass = {astro-ph.HE},
       adsurl = {https://ui.adsabs.harvard.edu/abs/2015PhRvL.114i5002L}
}

@ARTICLE{BesshoBhattacharjee2012,
       author = {{Bessho}, Naoki and {Bhattacharjee}, A.},
        title = "{Fast Magnetic Reconnection and Particle Acceleration in Relativistic Low-density Electron-Positron Plasmas without Guide Field}",
      journal = {\apj},
         year = 2012,
        month = may,
       volume = {750},
       number = {2},
          eid = {129},
        pages = {129},
          doi = {10.1088/0004-637X/750/2/129},
       adsurl = {https://ui.adsabs.harvard.edu/abs/2012ApJ...750..129B}
}

@article{Liu2015,
  title = {Scaling of Magnetic Reconnection in Relativistic Collisionless Pair Plasmas},
  author = {Liu, Yi-Hsin and Guo, Fan and Daughton, William and Li, Hui and Hesse, Michael},
  journal = {Phys. Rev. Lett.},
  volume = {114},
  issue = {9},
  pages = {095002},
  numpages = {5},
  year = {2015},
  month = {Mar},
  publisher = {American Physical Society},
  doi = {10.1103/PhysRevLett.114.095002},
  url = {https://link.aps.org/doi/10.1103/PhysRevLett.114.095002}
}

@ARTICLE{Guo2021,
       author = {{Guo}, Fan and {Li}, Xiaocan and {Daughton}, William and {Li}, Hui and {Kilian}, Patrick and {Liu}, Yi-Hsin and {Zhang}, Qile and {Zhang}, Haocheng},
        title = "{Magnetic Energy Release, Plasma Dynamics, and Particle Acceleration in Relativistic Turbulent Magnetic Reconnection}",
      journal = {\apj},
         year = 2021,
        month = oct,
       volume = {919},
       number = {2},
          eid = {111},
        pages = {111},
          doi = {10.3847/1538-4357/ac0918},
archivePrefix = {arXiv},
       eprint = {2008.02743},
 primaryClass = {astro-ph.HE},
       adsurl = {https://ui.adsabs.harvard.edu/abs/2021ApJ...919..111G}
}

@article{Mehlhaff2020,
    author = {Mehlhaff, J M and Werner, G R and Uzdensky, D A and Begelman, M C},
    title = {Kinetic beaming in radiative relativistic magnetic reconnection: a mechanism for rapid gamma-ray flares in jets},
    journal = {Monthly Notices of the Royal Astronomical Society},
    volume = {498},
    number = {1},
    pages = {799-820},
    year = {2020},
    month = {08},
    issn = {0035-8711},
    doi = {10.1093/mnras/staa2346},
    url = {https://doi.org/10.1093/mnras/staa2346},
    eprint = {https://academic.oup.com/mnras/article-pdf/498/1/799/33715054/staa2346.pdf},
}

@ARTICLE{Daughton2011,
       author = {{Daughton}, W. and {Roytershteyn}, V. and {Karimabadi}, H. and {Yin}, L. and {Albright}, B.~J. and {Bergen}, B. and {Bowers}, K.~J.},
        title = "{Role of electron physics in the development of turbulent magnetic reconnection in collisionless plasmas}",
      journal = {Nature Physics},
         year = 2011,
        month = jul,
       volume = {7},
       number = {7},
        pages = {539-542},
          doi = {10.1038/nphys1965},
       adsurl = {https://ui.adsabs.harvard.edu/abs/2011NatPh...7..539D}
}

@INPROCEEDINGS{Dahlin2015,
       author = {{Dahlin}, Joel Timothy and {Drake}, James and {Swisdak}, Marc},
        title = "{Electron Acceleration in 2D and 3D Magnetic Reconnection}",
    booktitle = {Solar Heliospheric and INterplanetary Environment (SHINE 2015)},
         year = 2015,
        month = jul,
          eid = {17},
        pages = {17},
       adsurl = {https://ui.adsabs.harvard.edu/abs/2015shin.confE..17D}
}

@ARTICLE{Bacchini2025,
       author = {{Bacchini}, Fabio and {Werner}, Gregory R. and {Granier}, Camille and {Vos}, Jesse},
        title = "{Three-dimensional Dynamics of Strongly Magnetized Ion{\textendash}Electron Relativistic Reconnection}",
      journal = {\apjl},
         year = 2025,
        month = sep,
       volume = {991},
       number = {1},
          eid = {L9},
        pages = {L9},
          doi = {10.3847/2041-8213/ae0197},
archivePrefix = {arXiv},
       eprint = {2507.12509},
 primaryClass = {astro-ph.HE},
       adsurl = {https://ui.adsabs.harvard.edu/abs/2025ApJ...991L...9B}
}

@INPROCEEDINGS{Dahlin2017,
       author = {{Swisdak}, M. and {Dahlin}, J.~T. and {Drake}, J.~F.},
        title = "{The role of three-dimensional transport in driving enhanced electron acceleration during magnetic reconnection}",
    booktitle = {AGU Fall Meeting Abstracts},
         year = 2017,
       series = {AGU Fall Meeting Abstracts},
       volume = {2017},
        month = dec,
          eid = {SH12B-03},
        pages = {SH12B-03},
       adsurl = {https://ui.adsabs.harvard.edu/abs/2017AGUFMSH12B..03S}
}

@article{Petropoulou2018,
    author = {Petropoulou, Maria and Sironi, Lorenzo},
    title = {The steady growth of the high-energy spectral cut-off in relativistic magnetic reconnection},
    journal = {Monthly Notices of the Royal Astronomical Society},
    volume = {481},
    number = {4},
    pages = {5687-5701},
    year = {2018},
    month = {10},
    issn = {0035-8711},
    doi = {10.1093/mnras/sty2702},
    url = {https://doi.org/10.1093/mnras/sty2702},
    eprint = {https://academic.oup.com/mnras/article-pdf/481/4/5687/26762394/sty2702.pdf},
}

@article{Cerutti2014,
doi = {10.1088/0004-637X/782/2/104},
url = {https://doi.org/10.1088/0004-637X/782/2/104},
year = {2014},
month = {feb},
publisher = {The American Astronomical Society},
volume = {782},
number = {2},
pages = {104},
author = {Cerutti, B. and Werner, G. R. and Uzdensky, D. A. and Begelman, M. C.},
title = {THREE-DIMENSIONAL RELATIVISTIC PAIR PLASMA RECONNECTION WITH RADIATIVE FEEDBACK IN THE CRAB NEBULA},
journal = {The Astrophysical Journal}
}

@ARTICLE{Parker1963,
       author = {{Parker}, E.~N.},
        title = "{The Solar-Flare Phenomenon and the Theory of Reconnection and Annihiliation of Magnetic Fields.}",
      journal = {\apjs},
         year = 1963,
        month = jul,
       volume = {8},
        pages = {177},
          doi = {10.1086/190087},
       adsurl = {https://ui.adsabs.harvard.edu/abs/1963ApJS....8..177P}
}

@article{Borgogno2005,
    author = {Borgogno, D. and Grasso, D. and Porcelli, F. and Califano, F. and Pegoraro, F. and Farina, D.},
    title = {Aspects of three-dimensional magnetic reconnection},
    journal = {Physics of Plasmas},
    volume = {12},
    number = {3},
    pages = {032309},
    year = {2005},
    month = {02},
    issn = {1070-664X},
    doi = {10.1063/1.1857912},
    url = {https://doi.org/10.1063/1.1857912},
    eprint = {https://pubs.aip.org/aip/pop/article-pdf/doi/10.1063/1.1857912/14722385/032309_1_online.pdf},
}

@article{Baalrud2018,
    author = {Baalrud, S. D. and Bhattacharjee, A. and Daughton, W.},
    title = {Collisionless kinetic theory of oblique tearing instabilities},
    journal = {Physics of Plasmas},
    volume = {25},
    number = {2},
    pages = {022115},
    year = {2018},
    month = {02},
    issn = {1070-664X},
    doi = {10.1063/1.5020777},
    url = {https://doi.org/10.1063/1.5020777},
    eprint = {https://pubs.aip.org/aip/pop/article-pdf/doi/10.1063/1.5020777/13396421/022115_1_online.pdf},
}

@article{Granier2024,
    author = {Granier, C. and Tassi, E. and Laveder, D. and Passot, T. and Sulem, P. L.},
    title = {Influence of ion-to-electron temperature ratio on tearing instability and resulting subion-scale turbulence in a low-βe collisionless plasma},
    journal = {Physics of Plasmas},
    volume = {31},
    number = {3},
    pages = {032115},
    year = {2024},
    month = {03},
    issn = {1070-664X},
    doi = {10.1063/5.0185897},
    url = {https://doi.org/10.1063/5.0185897},
    eprint = {https://pubs.aip.org/aip/pop/article-pdf/doi/10.1063/5.0185897/19844154/032115_1_5.0185897.pdf},
}

@article{TenBarge2013,
doi = {10.1088/2041-8205/771/2/L27},
url = {https://doi.org/10.1088/2041-8205/771/2/L27},
year = {2013},
month = {jun},
publisher = {The American Astronomical Society},
volume = {771},
number = {2},
pages = {L27},
author = {TenBarge, J. M. and Howes, G. G.},
title = {CURRENT SHEETS AND COLLISIONLESS DAMPING IN KINETIC PLASMA TURBULENCE},
journal = {The Astrophysical Journal Letters}
}

@article{Zhang2023,
doi = {10.3847/2041-8213/acfe7c},
url = {https://doi.org/10.3847/2041-8213/acfe7c},
year = {2023},
month = {oct},
publisher = {The American Astronomical Society},
volume = {956},
number = {2},
pages = {L36},
author = {Zhang, Hao and Sironi, Lorenzo and Giannios, Dimitrios and Petropoulou, Maria},
title = {The Origin of Power-law Spectra in Relativistic Magnetic Reconnection},
journal = {The Astrophysical Journal Letters}
}
\bibliographystyle{aasjournal}

\end{document}